\documentstyle[aps,twocolumn,prb,epsf]{revtex}
\renewcommand{\ref}[1]{\raisebox{.6ex}{[#1]}}

\newcommand{\be}{\begin{equation}}
\newcommand{\ee}{\end{equation}}

\newcommand{\ba}{\begin{array}}
\newcommand{\ea}{\end{array}}

\newcommand{\bea}{\begin{eqnarray}}
\newcommand{\eea}{\end{eqnarray}}

\begin{document}


\twocolumn[\hsize\textwidth\columnwidth\hsize\csname @twocolumnfalse\endcsname

\title{ Microscopic Theory of Vortex Dynamics in Homogeneous Superconductors }

\author{ P. Ao \\ Department of Theoretical Physics  \\
     Ume\aa{\ }University, S-901 87, Ume\aa, SWEDEN   \\
        and \\
      X.-M. Zhu \\ Department of Experimental Physics   \\
           Ume\aa{\ }University, S-901 87, Ume\aa, SWEDEN}

\maketitle

\widetext

\begin{abstract}
Vortex dynamics in a type-II superconductor 
is systematically investigated by the influence functional method. 
The irrelevant fermionic degrees of freedom  are integrated out and
their effects on the dynamics are treated in terms of the vortex coordinate.
When an isolated vortex is moving against its background, 
forces proportional to the first order of vortex velocity
on the vortex are calculated within the present formulation. 
The total transverse force on the moving vortex
is explicitly shown to be proportional to the superfluid number density
and insensitive to impurities.
Its equivalent expressions in terms of the Berry phase and  
the various summations of transitions between quasiparticle (hole) 
states are discussed.
At finite temperatures, due to the finite population of quasiparticle (hole)
excitations above (below) the energy gap, 
there is a friction against vortex motion which
diverges logarithmically in the low-frequency limit.
Nonmagnetic impurities give rise to an additional friction
from the core states 
which saturates to a value independent of
the  normal state resistivity in the dirty limit.
In this limit, 
the coupling to  the electromagnetic field does not change the
conclusions if charge neutrality  in the superconductor is maintained. 
Macroscopic constraints on vortex dynamics   by the 
second law of thermodynamics and by the fluctuation-dissipation theorems 
are also discussed.

\end{abstract}

\pacs{ 74.60.Ge; 67.40.Vs }

]



\narrowtext

\section{Introduction } 


In a type-II superconductor, vortex motion is responsible for a variety of 
low frequency transport phenomena. 
It is the only topological singularity
whose dynamical properties are widely accessible to experimental studies
in both classical and quantum regimes, and 
its importance  has long been realized.\cite{bs,nv,tinkham,thouless} 
Despite decades of research, the
theoretical agreement reached so far is very limited:
At zero temperature, in the absence of any impurity potentials,
a vortex follows the local superfluid velocity. 
In the absence of a local superfluid velocity, when a vortex follows
the motion of an external trapping potential,
there is a momentum change in the superfluid 
transverse to the direction of vortex motion. 
In order to provide this momentum change, a force must be applied by the
external trapping potential to the superfluid through the vortex. 
The vortex experiences a transverse force proportional 
to the superfluid number density,
balanced by the external force from the trapping potential.
Beyond this simplest and idealized situation
many aspects of vortex dynamics have remained unsettled and even controversial.
In the present paper we attempt to provide an influence functional formulation
of vortex dynamics from the microscopic Bardeen-Cooper-Schrieffer (BCS) theory,
and a few detailed microscopic calculations under realistic conditions.

The current microscopic understandings of vortex dynamics
in the presence of impurities and at finite temperatures
may be classified into two different physical pictures,  
which are based on different theoretical approaches 
and give contradictory results.
In one picture the magnitude of the total force 
experienced by the vortex in the transverse direction is 
proportional to the superfluid number density.\cite{at,tan}
The superfluid momentum change caused by the
vortex motion is provided by an externally controlled trapping potential
in the absence of a local superfluid velocity,
regardless of the existence of the normal fluid.\cite{tan} 
It is an exact consequence of 
the global topological constraint on the vortex. 
The normal fluid at finite temperatures
gives rise to friction for the vortex motion in the longitudinal direction.
Furthermore, the global methods used in Refs.\ \onlinecite{at} and
\onlinecite{tan}
indicate that the total transverse force is insensitive to random impurities,
though there are additional frictional effects.
In this picture, 
in the absence of the externally controlled trapping potential, 
the pinning and friction should be used to obtain vortex motion perpendicular
to the direction of an externally applied current.  
For the other picture, the essence of the results is that
there are additional forces 
in the transverse direction of vortex velocity,
provided by unbounded quasiparticle
excitations or the normal fluid, by bounded vortex core states,  
by the substrate, or by a certain combination of 
them.\cite{volovik,simanek,kopnin,otterlo,stone0} 
The total transverse force is reduced, which is most clearly represented by
the alleged gradual turning on the cancelation 
between two topological effects by a relaxation time:\cite{volovik}
the spectral flow of vortex core state transitions
and the Berry phase counting far away from the  core. 
To discuss this controversy from a detailed and straightforward 
approach is one of the main purposes of the present paper. 

We now state precisely 
the physical quantities which we are going to address.
In the classical limit, we are looking for an effective equation of
motion for a vortex. In two dimensions (2D), 
or for a straight vortex line in 3D, 
the equation for a vortex specified by ${\bf r}_v$ 
takes the form of a Langevin equation:
\be
  m_v \ddot{\bf r}_v = {\bf F}({\bf r}_v,t) 
             - B \; \dot{\bf r}_v \times \hat{ z} 
             - \eta  \; \dot{\bf r}_v + {\bf f} \; .
\ee
Here $m_v$ is the vortex mass, $\eta$ is the friction coefficient,  $B$ is the 
coefficient for the transverse force
with $\hat{ z}$ the direction perpendicular to the 
plane of vortex motion, 
and the fluctuating force $\bf f$ related to the friction force by the 
fluctuation-dissipation theorem.
The force ${\bf F}$  contains all other forces which are not functions
of vortex velocity: the force from the trapping potential, pinnings, 
the force due to an externally applied supercurrent,\cite{wexler} 
the force due to other vortices, etc.
We may classify the terms in the above equation into three types,
according to the order of time derivatives of ${\bf r}_v$: 

{\bf i}). 
 Forces contain no explicit dependence on any time derivative of 
 vortex coordinate ${\bf r}_v $,  represented by ${\bf F}$. 
 These types of forces may be regarded as conceptually well understood,
 corresponding to the Born-Oppenheimer potentials, 
 and are not controversial, though practically 
 they can be difficult to evaluate.
 They are contained in our formulation,
 but will not be discussed in the present paper. 

{\bf ii}). 
 Forces have a first-order time derivative of the vortex coordinate, 
 the vortex velocity, 
 represented by the transverse and longitudinal coefficients $B$ and $\eta$.
 Calculating those forces explicitly is the focus of the present paper. 
 It is our purpose to clarify the physical origins behind those forces, 
 starting from a well-defined microscopic theory, the BCS theory, 
 using a well-defined and rigorous procedure, the influence functional 
 method.\cite{leggett}  
 We will show in detail that the total transverse force 
 is insensitive to details, and is proportional to the superfluid number 
 density, and present calculations leading 
 to finite vortex friction contributions. 

{\bf iii}). 
   The term contains the second-order 
   time derivative of the vortex coordinate, the vortex mass $m_v$.
   This is also an unclear quantity, and the subject of the recent active 
   study.\cite{niu,gr} Though we believe our present formulation
   also  provides the framework to address the dynamical effects
   on the vortex mass, 
   we will not explore them here.\cite{han}
   This term will be ignored by
   assigning the vortex acceleration $ \ddot{\bf r}_v = 0 $.

We organize the rest of the paper as follows.
In Sec. II the total transverse force 
is studied from a macroscopic point of view.
We first demonstrate from a thermodynamic consideration that the magnitude 
of the total transverse force should be proportional to the superfluid number 
density. A reduction from this value
will lead to a violation of the second law of thermodynamics.
Then we put the transverse force and friction in the context of  
fluctuation-dissipation theorems, and illustrate that the relaxation
time approximation in the microscopic derivations of 
vortex dynamics should be avoided.
In Sec. III we first present a general formulation
based on the BCS theory. Then we relate this formulation to that of the
influence functional approach which has been proved to be rigorous and
effective to calculate friction in quantum dissipative dynamics
of a subsystem, where the total system is described by a Hamiltonian. 
A few general properties of our formulation will be discussed.
In Sec. IV 
we give detailed evaluations of both longitudinal and transverse
correlations in the clean limit for arbitrary temperatures,
which lead to both the friction and the total transverse force.
In particular,  detailed evaluations of the 
total transverse force from either extended state counting 
or core state transitions are given there, 
and are explicitly shown to be equivalent.  
In Sec. V  the effects of impurity potentials are considered.
We  will show that the total transverse force will not be affected.
However, impurity potentials strongly affect the core state spectrum, which
leads to a contribution to the vortex friction
in addition to that of  extended states.
In Sec. VI we show that the electromagnetic field does not affect
the total transverse force and friction, under the condition that 
the charge neutrality in the superconductor is maintained.
In Sec. VII some experimental tests are briefly discussed, and
we summarize in Sec. VIII.

\section{ Thermodynamics and Statistical Mechanics }

\subsection{ Force Balance and Thermodynamics }

The microscopic calculations which 
we will present later are unavoidably lengthy and technical.
It may be  helpful to obtain an overall picture and useful 
information (as much as we can) under general but elementary considerations.
In this subsection we give a thermodynamic consideration 
of vortex dynamics to show that there is a constraint on the 
total transverse force,
and in the next subsection the derivation of vortex dynamics will be put in
the context of fluctuation-dissipation theorems.

We may write down a possible
equation of motion for a quantized vortex
in the absence of impurities, taking into account
the possible role of the normal fluid in the motion of the  vortex,
in the limit of the vortex acceleration equal to zero: 
\[
   h \rho_s\hat{z}\times({\bf u}_v - {\bf v}_s)
   - D   ( {\bf u}_v - {\bf v}_n )
   - D'  \hat{z} \times ( {\bf u}_v - {\bf v}_n )
\]
\be
    + \; {\bf F}^{ext} = 0  \, . 
\ee
Here $ \rho_s$ is the number density of the superfluid,
${\bf u}_v $,  ${\bf v}_s$ and ${\bf v}_n$ are the 
 velocities of vortex, the superfluid,  and the normal fluid 
with respect to the substrate or the wall of the container.
Those velocities are independent variables. 
The velocity dependences are only in first order in Eq. (2). 
The first term in the left-hand side of Eq. (2) is the Magnus force,\cite{nv} 
whose magnitude is proportional to the superfluid number density. 
The last term represents a possible external force on the vortex. 
The other two terms are possible contributions coming from the
interaction of the vortex with the normal fluid.
Initially, both the normal fluid and superfluid velocities are set to zero.

We will demonstrate that the conditions of force balance
and thermodynamics put a constraint on the value of $ D' $.
For this purpose let us imagine a torus-shaped tank filled with a superfluid,
or a torus-shaped superconductor film. 
The tank can be considered as a thermal reservoir to the superfluid.
This implies that at finite temperatures there is also a normal fluid.
After creating a vortex-antivortex pair, we keep the antivortex at rest
and move the vortex to wind once with a small velocity $u_v$ 
around one of the two circumferences of the torus, say $L_y$,
in time $t_{total}$ before the annihilation with the antivortex.
We take $t_{total} $ much longer than the relaxation time 
of the normal fluid
such that the normal fluid velocity always stays close to zero, 
by transferring a possible momentum gained from the vortex motion
to the substrate, via the relaxation process represented
by the normal fluid viscosity. 
A physical realization may be the electron-phonon interaction.
Hence, the normal fluid velocity is always negligible comparing with
the vortex velocity ${\bf u}_v$ 
which is an order of $u_v \approx L_y/t_{total}$. 
As a result of the vortex motion,  the momentum of circulating superfluid
particles along the torus has been changed from zero to $p_{s,f} = h/ L_x$,
because of the change of the winding number of the superfluid.
This occurs regardless how slow the vortex motion is.
The kinetic energy of the superfluid has been changed 
from zero to $ E = \rho_s L_x L_y p_{s,f}^2 /2 m^{\ast} $ for 
a neutral superfluid, or when the effective 
magnetic screening length is larger than $L_y$ for the superconductor film.
Here  $m^{\ast}$ is the effective mass of superfluid particles, 
and $L_x$ and $L_y$ are the circumferences of the torus. 

The total momentum change of the superfluid requires a 
force in the transverse direction of the vortex motion
\[
  F_{\perp} = \frac{dP}{dt}
            = \frac{ \rho_s  L_x L_y \, p_{s,f} } {t_{total} }
            = h \rho_s u_v  \; ,
\]
here $P$ is the total momentum of the superfluid.
Since the normal fluid velocity stays zero, 
no kinetic energy can be transferred from the normal fluid to the superfluid. 
However, if the normal fluid would contribute to this force to superfluid
by changing its internal energy, an
additional transverse force on the vortex, 
$-D' \hat{z} \times {\bf u}_v$,  arises. 
The magnitude of the 
external force in the transverse direction of vortex motion should be
equal to the total transverse force  according to Eq. (2),
\[
  F_{\perp}^{ext}
  = ( h \rho_s -D'  )u_v \, .
\]

Now we are ready to consider the thermodynamic relations.
The process of creating a vortex-antivortex pair and its annihilation
after the vortex crossing one circumference $L_y$ of the torus  
leaves only a finite increase of superfluid circulation in the tank,
corresponding to the change of winding number. 
The initial and final normal fluid velocities are zero.
The increase of kinetic energy of the superfluid 
needs to be provided from somewhere.
There are only two possible sources: the external 
trapping potential and the normal fluid.
Here we need to be reminded of 
 a significant difference between 
the superfluid and the normal fluid: The superfluid carries no entropy, 
while the normal fluid does.
Therefore, according to the second law of thermodynamics,\cite{callen} the
superfluid cannot gain kinetic energy by lowering
the internal energy or entropy of the normal fluid.

We need to consider the work performed on the system by the
external force to move this vortex. 
In the longitudinal direction of vortex motion,
the interaction between the normal fluid and the vortex
gives rise to a vortex friction $ -D u_v$.
Thus the  external force  on the normal fluid
in the longitudinal direction  is $ D u_v $.
This friction does not dissipate energy.
Rephrased alternatively,
the energy dissipated can be arbitrarily small by taking the time 
to complete the process arbitrarily long,
$t_{total}\rightarrow\infty$.
The process is then quasistatic.
The  normal fluid velocity is always negligible in the process
because of its finite viscosity.
Thus  the external force  on the
normal fluid in the longitudinal direction of vortex motion
does not provide any work to the system.
The external force acting on the superfluid will be
able to provide enough work for the kinetic 
energy increase only if  $F_{\perp}^{ext} \geq F_{\perp} $, 
{\it i.e.}, $D' \leq 0$, which leads to the conclusion
that the magnitude of total transverse force cannot be reduced
from that determined by the superfluid number density.
The  work done by the external force 
is  exactly equal to the kinetic energy increase  
if the magnitude of the external  force  in the transverse direction is
the product of the superfluid number density, the Planck constant $h$, 
and the vortex velocity. 

The next question is whether or not the total transverse
force on a moving vortex can be larger than that determined by the 
superfluid number density. 
If the normal fluid would carry a vortex with a 
vorticity in the same direction as that of the superfluid,
the answer to this question is positive.
However, since we have assumed that the normal fluid is
viscous, the vortex of the normal fluid will eventually disappear.
This is true for a slow process whose time scale is much larger
than the relaxation time of the normal fluid assumed here.
This consideration leads to that the
total transverse force cannot be larger than the value determined by the
superfluid number density.
Combining with the thermodynamic argument we conclude $ D' = 0 $. 

The above discussion has explicitly made use of the assumption of a  
finite normal fluid viscosity.
In case that the normal fluid relaxation time would be infinite, that is, 
the normal fluid viscosity would be zero, 
a process which generates a vortex circulation
in the superfluid would also generate a vortex circulation in the normal fluid.
This would be the limiting situation of a dynamical process 
in which the internal relaxation time of the normal fluid is much shorter 
than its relaxation time to the substrate and 
the time scale for the process is between them.
An example would be the creation of vortices by 
a magnetic flux in an ultraclean  superconductor. 
In such a case, the normal fluid velocity will not relax to zero.
Under this ideal condition $ - D'  = h \rho_n$, 
corresponding to that the normal fluid  has a vortex, which
is what has been discussed in Ref.\onlinecite{gwt}.
   
If impurities are present, a phenomenological
equation of motion for the vortex may be written down if
the impurities are homogeneously distributed and vary only
at a scale much smaller compare with the size of the vortex core.
We have two more possible parameters from the vortex-impurity
interaction:
\[
  h \rho_s\hat{z}\times({\bf u}_v - {\bf v}_s)
  - D  ({\bf u}_v - {\bf v}_n )
  - D' \hat{z}\times ( {\bf u}_v - {\bf v}_n ) 
\]
\be 
   -d \, {\bf u}_v - d'\hat{z}\times {\bf u}_v + {\bf F}^{ext} = 0 \; .
\ee
Parallel to what we have discussed for the normal fluid
case, $ d'$ must be zero in order for the external force
to provide the energy gain needed by the superfluid. The
impurities cannot provide energy to the superfluid either by 
lowering their internal energy or entropy
because of the second law of thermodynamics.
We note that impurities introduce another contribution to 
the normal fluid viscosity.

The microscopic global considerations\cite{at,tan} have already suggested  
$D'=0$ and $d'=0$. 
The conclusion here will be borne out by 
detailed and independent microscopic calculations in the following sections.

\subsection{ Friction and Fluctuation-Dissipation Theorems }

The derivation of the equation of motion for the vortex is
different from the usual linear-response theory. 
In the linear-response theory, a driving force
is given, that is, the Hamiltonian is known, and
we look for the average responding velocity.
It is a calculation of conductivity or mobility.
In the present case the force on a moving vortex is 
the unknown quantity which we need to find out.
The vortex velocity is, however, readily defined 
through the vortex coordinate.
It is a calculation of resistivity or friction.
To appreciate this difference, 
we will examine the different correlation functions
involved in these different types of calculations and their relationships.  
The focus point in this subsection 
in on the condition for using the relaxation time approximation.

When the normal fluid is at rest, 
the vortex motion is governed by a classical Langevin equation with 
parameters to be determined microscopically.  
This equation has the same form as a classical electron moving in a
magnetic field.  We adopt the language in transport theory 
to make it easier to relate to the early  work in that 
field.\cite{kubo,bh,hc,fm}

We start by considering a classical charged particle in a magnetic
field obeying a generalized
Langevin equation:
\bea
   m \dot{u}_i(t) & = &  -\int^{t}_{t_0} dt' \; \eta_{ik}(t-t')  u_k(t')
        + { F}^{ext}_i(t)  \nonumber  \\
      & & - B \epsilon_{ik} \; u_k(t) + f_i(t) \; .
\eea
Here $i = x$ or $y$,
${\bf u}(t) = ( u_x(t), u_y(t) ) $ is the velocity of the particle,
 $m$ is
the mass, ${\bf F}^{ext}(t)= ( F^{ext}_x(t), F^{ext}_y(t) ) $ 
is an external force, ${\bf f}(t) = ( f_x(t),  f_y(t) )$ is a random
force which simulates the effect of the thermal reservoir.
The Einstein convention of the repeated indices as summation has been used.
$B \epsilon_{ik}u_k(t)$
represents the transverse force, the Lorentz force
$ - {\bf u}(t) \times {\bf B}$ in the Langevin equation with
the magnetic field taken along the $z$-direction. 
The matrix $\eta(t-t') = \{ \eta_{ij} \} $ represents friction in both 
longitudinal and transverse directions of the particle motion.
Its possible finite off-diagonal elements
will change the effect of the original Lorentz force on the particle.
In addition, we have 
\bea
   < f_i (t)> & =  & 0 \; ,   \nonumber  \\
   <  u_i (t_0) f_j (t_0 + t )> & = & 0 \; , \; \, t > 0        \; ,  \\
   <u_i(t_0) u_j(t_0) > & =  &  \frac{k_B T}{m} \delta_{ij} \; .  
    \nonumber   
\eea
The first equation is obvious: no average fluctuating force.
The second one is due to the causality and the
last one is due to the equipartition theorem.

If the Hamiltonian of the particle is known, 
the problem of particle responding to a perturbation
can  be formulated in two different but equivalent ways. 
We can calculate the velocity of the particle while the
applied force is given. 
In such a case, it is to obtain a conductivity or mobility formula. 
The conductivity or mobility may be  obtained by the Nakano-Kubo's formula, 
a calculation of velocity-velocity correlation function.
It may also be obtained by solving the Boltzmann equation
in the presence of an electric field.\cite{hc}
Otherwise, we can consider a given velocity for the particle and
calculate the applied force needed to maintain this motion.
It is to
obtain a resistivity or friction formula, i.e., calculating an electric
field needed to maintain the given current.
The derivation of vortex dynamics  belongs to the
second kind, where we consider a steady motion of the vortex and calculate
the external force acted on the vortex.
Unfortunately, we do not have the choice to formulate vortex dynamics
in superconductors
in terms of conductivity or mobility formula because the effective 
vortex Hamiltonian is unknown {\it a priori}.

Introducing a Laplace transform
\[
  \eta[\omega] = \int^{\infty}_{0}  dt \; e^{-i\omega t}\eta(t)  \, ,
\]
the mobility  is given from Eq. (4) in the limit $t_0\rightarrow -\infty$ by
\be
  \mu[\omega] = (im\omega + \eta[\omega] + i \sigma_y B )^{-1} \: .
\ee
Here the mobility $\mu[\omega]$ is defined through
\[
  <{\bf u}_i [\omega]> = \mu_{ij}[\omega] 
        \overline{\bf F}^{ext}_{j}[\omega] \; ,
\]
with an applied external force
 $\overline{\bf F}^{ext}(t) =\overline{\bf F}^{ext}[\omega] e^{i\omega t} $. 

Defining the velocity-velocity correlation function matrix
\[
   {\cal U}_{ij}(t) =  < u_i(t_0 + t) u_j(t_0) > \; ,  
\]
with ${\cal U}_{ij}(t=0) = \delta_{ij} k_B T/m$ according to Eq. (5),
the mobility is related to the velocity-velocity 
correlation function 
\be
  \mu[\omega] = \frac{{\cal U}[\omega] }{k_B T} \, .
\ee 
This is the `first' fluctuation-dissipation theorem described by 
Kubo,\cite{kubo}
equivalent to the Nakano-Kubo's formula for the electrical conductivity. 

It is easy to demonstrate that the relaxation time approximation
can be valid in the Nakano-Kubo's formula. 
Without the thermal reservoir, the velocity-velocity
correlation is given by
\be
   {\cal U}[\omega] = (i m \omega + i \sigma_y B )^{-1}  
     m \; {\cal U}(0) \; .
\ee
When using a relaxation time
approximation by the standard rule,
$i\omega\rightarrow i\omega + \eta [\omega]/m$ and
substituting it into Eq. (8), 
we find the velocity-velocity
correlation under the relaxation time approximation is given by
\[
  {\cal U}[\omega]
  = (i m \omega + \eta[\omega] + i \sigma_y B )^{-1}  k_B T\; ,
\]
which is exactly the same as the one obtained by  the rigorous
calculation, Eqs. (6) and (7).
Therefore, the relaxation time approximation can be a valid one
for velocity-velocity correlations when used in a conductivity 
or mobility formula.

The resistivity or friction formula 
is known to be difficult and it is worthwhile to
examine it closely.\cite{kubo,hc,bh,fm}
First, we calculate the total force-force correlation function matrix
\[
  {\cal F}_{ij}(t) = m^2 <\dot{u}_i (t_0 + t) \dot{u}_j (t_0)> \, .
\]
Taking the Laplace transform, using the 
translational invariance in time
\[
  < u_i (t_0+t) \dot{u}_j (t_0)> = -<\dot{u}_i (t_0+t) u_j(t_0)> \; ,
\]
and the total force-velocity correlation function
\[ 
  m <\dot{u}_i (t_0+t) u_j(t_0)>[\omega] = - m {\cal U}_{ij}(0)
     + i m \omega {\cal U}_{ij}[\omega ] \; ,
\]
we have
\be
  {\cal F} [\omega]  =  \left( - iB\sigma_y + i m \omega +
  \frac{ (m\omega)^2}
       { im \omega + \eta [\omega] + i \sigma_y B  } \right) k_B T  \, . 
\ee
In the limit $\omega\rightarrow 0$, it is reduced to
\be
  {\cal F} [0] = - i B \sigma_y \; k_B T  \; ,
\ee
and is independent of the frictional coefficient $\eta$. 

The random force-force correlation matrix is defined as
\be
   {\cal R}_{ij}(t) = <f_i (t_0 + t) f_j(t_0)> \; . 
\ee
>From the Langevin equation, Eq. (4), 
we can express  ${\cal R}(t)$ in terms of total force-force, 
total force-velocity, and velocity-velocity correlation functions.
Taking the Laplace transform and integrating by part, 
we obtain
\be
   {\cal R}[ \omega ] 
   = \eta[\omega] \; m {\cal U}(0) =  \eta[\omega] \; k_B T \, ,
\ee
or $\eta(t) =  {\cal R}(t) /(k_B T ) $.
This is the `second' fluctuation-dissipation theorem described by 
Kubo.\cite{kubo} 
We emphasize that the generalized frictional coefficient
$\eta(t)$ is directly given by the random force-force correlation. 
The frictional coefficient matrix
$\eta(t)$ has no off-diagonal part if the random force-force
correlation matrix has not.
This fluctuation-dissipation theorem allows us to 
obtain some general properties of the generalized friction.
For example, for a charged particle 
described by a single relaxation time in the Boltzmann equation
moving in a magnetic field, there will be no frictional effect 
on the force in the transverse direction.

Next we consider that the particle is moving at a given 
velocity $\overline{\bf u}(t)$ and find out what is the 
external force needed to sustain such a motion. This is exactly the
situation which we encounter in vortex dynamics and
it is equivalent to the calculation of
resistivity or friction. 
>From Eq. (4),  the average force is given by
\be
   < F^{ext}_i[\omega] > 
  = (im \omega + \eta[\omega] + i \sigma_y B )_{ij}
          \; \overline{u}_{j}[\omega]  \; ,
\ee
which is trivially identical to the reciprocal of
conductivity formula. Obviously, this process does not provide us an
independent way of calculating resistivity.

However, if we are only interested in the average force  
$ <{\bf F}^{ext} > $ in the DC limit,  
we do have an alternative resistivity formula. 
After  taking $ \omega\rightarrow 0$ and using Eqs. (10) and (12),
Eq. (13) gives
\be 
   < F^{ext}_i >[0]
    = \frac{ 1 }{k_B T}  \left( {\cal R}_{ij}[0]
         - {\cal F}_{ij}[0]  \right) \overline{u}_j[ 0 ]  \, .
\ee
Taking  $\eta$ to be a scalar, 
the external force can be written into  a more suggestive form,
\be
   <{\bf F}^{ext} [0]> = \eta[0] \; \overline{\bf u}[0] 
                + \overline{\bf u}[0]\times {\bf B} \; ,
\ee
where the longitudinal component depends on ${\cal R}[0]$, 
the random force-force correlation function, 
and the transverse component only on ${\cal F}[0]$, the total force-force 
correlation function.  
Equation (14) is the DC resistivity formula. 
It provides a direct way to obtain
DC resistivity from force correlation functions.  
The straightforward interpretation of Eq. (15) is the force-balance: 
The externally applied force to keep the constant velocity
is equal in magnitude but opposite in sign 
to the sum of the frictional and the Lorentz forces.

We will show that the relaxation time approximation
is invalid when used in the force calculation.
We start with the force-force correlations.
Without  thermal reservoir, 
the random force correlation is zero, that is, ${\cal R}(t) = 0$.
If we switch on the effect of a thermal  reservoir by using a relaxation time 
approximation $  i\omega \rightarrow i\omega + \eta[\omega]/m$,
the random force correlation is still
incorrectly set to zero. This shows that the relaxation time approximation
cannot be used to calculate the  random force correlation.

The total force correlation without thermal reservoir is
\be
    {\cal F}_{ij} [\omega]  =  
  \left( - i B\sigma_y  + i m \omega 
   + \frac{ (m\omega)^2 }{ i m \omega  + i \sigma_y B }
        \right)_{ik} m {\cal U}_{kj}(0)  \, .
\ee
When we switch on the thermal  reservoir by using a relaxation time 
approximation $  i\omega \rightarrow i\omega + \eta[\omega]/m$ 
in Eq. (16), we have 
\[
  {\cal F} [\omega]  = 
   \left(- i B\sigma_y + i m \omega + \eta[\omega]
   - \frac{ (im \omega + \eta[\omega])^2 }
          { i m \omega  +   \eta[\omega] + i \sigma_y B  }
      \right) k_B T  \,  .
\]
This is a rather complicated expression. 
In the limit $\omega<< \eta[\omega]$, or $ \omega \tau<<1 $, 
we can simplify it to 
\be
   {\cal F}[0]
    = \frac{ B }{ 1+ (\omega_0 \tau)^2 } \left(
      \omega_0 \tau - i\sigma_y (\omega_0 \tau)^2 \right) k_B T \; , 
\ee
with 
$\omega_0 = B/m$. Here $\tau=m/\eta[0] $ is a relaxation time.
Let us use the resistivity formula Eq. (14) to calculate the external
force needed to keep the particle moving with a given 
velocity. With ${\cal R}[\omega] = 0$ and  ${\cal F}[0]$ given by Eq. (17),
the external force is
\be
   <{\bf F}^{ext} [0]> =   
     \frac{\omega_0 \tau}{ 1+(\omega_0 \tau)^2} 
     ( - B \overline{\bf u}[0] +   
       \omega_0\tau \; \overline{\bf u}[0] \times {\bf B} )  \, . 
\ee
These results have no connection at all to the rigorous
results shown in Eq. (15). 
Even the sign for the longitudinal force is wrong.
Evidently, the relaxation time
approximation cannot be valid in such a calculation,
because such an approximation in the force balance equation
cannot consider the random force properly, and leads to results
violating the fluctuation-dissipation theorems.

By the simple and exactly solvable model,
we have demonstrated the essential conditions for 
the validity of the relaxation time 
approximation in velocity-velocity correlation function calculations,
and for its invalidity in  force-force correlation function calculations.
We refer readers to Refs.\onlinecite{kubo,hc,bh,fm}
for more sophisticated discussions in the context of the  Green's
function or Boltzmann equation.

With  a redefinition of constants $B= h \rho_s$ with
$ h $ being the Planck constant,
$\rho_s$ being the superfluid particle number density, 
and $\omega_0 = \epsilon_0 $ as the core-level spacing,
this force becomes  the same in magnitude as 
the one which appeared in the derivation of vortex dynamics using the  
relaxation time approximation,\cite{volovik,simanek,kopnin,stone0} including  
the same sign error.\cite{otterlo} 
The inappropriate use of the relaxation time approximation in
vortex dynamics in $d$-wave superconductors has also been pointed 
out recently.\cite{ao}
In the following 
we show how to obtain the vortex friction without  
the relaxation time approximation, and demonstrate at the same time
that the total 
transverse force remains unchanged as dictated by the topology.

\section{ Vortex Dynamics in Homogeneous BCS Superfluid }

\subsection{ Formulation of the Problem } 

We present now our microscopic derivation, 
from the standard BCS Lagrangian for $s$-wave pairing in the imaginary time
path-integral formulation of the influence functional method. 
The connection of the total transverse force to the Berry phase
is straightforward in this formulation.
We believe the present formulation has some advantages:
a transparent crossover from the quantum  
to the classical description via the semiclassical approximation, and 
a flexible treatment of the general dissipative effect arising
from the integration out of irrelevant degrees of freedom, 
fermionic quasiparticles, and holes.
The relevant degree of freedom is the vortex coordinate.\cite{az}

We consider a neutral fermionic superfluid first.
The coupling to electromagnetic fields will be discussed later.   
The Lagrangian is given by
\bea
  L_{BCS} & = &  \sum_{\sigma}\psi^{\dag}_{\sigma}( x,\tau) 
        \left(  \hbar\partial_{\tau}- \mu_F - 
        \frac{\hbar^{2}}{2m} \nabla^{2} + V(x)  \right. \nonumber \\
      & & \left.  + U_0(x-x_v) \right) \psi_{\sigma}(x,\tau)   \nonumber \\
      & & - g\psi^{\dag}_{\uparrow}(x,\tau) \psi^{\dag}_{\downarrow}(x,\tau)
          \psi_{\downarrow}(x,\tau) \psi_{\uparrow}(x,\tau ) \; ,
\eea
where $\psi_{\sigma}$ describes electrons with spin $\sigma=(
\uparrow, \downarrow )$, $\mu_F$ is the chemical potential determined by the 
electron number density, $V(x) $ is the impurity potential, $ U_0 $ is
the trapping potential, and $x =(x,y,z)$.
A vortex at $x_v$ has been assumed through the trapping potential. 
A more explicit implementation of the vortex coordinate will be discussed 
after Eq. (39).  The partition function is 
\be
   Z = \int {\cal D}\{x_v, \psi^{\dag}, \psi \}              
           \exp\left\{ - \frac{1}{\hbar} \int_0^{\hbar\beta } 
             d\tau \int d^3x L_{BCS} \right\}   \; ,      
\ee
with $\beta = 1/k_B T $ , and $d^3x = dxdydz$.
Inserting the identity in the functional space,
\[ 
     1 =   \int {\cal D}\{ \Delta^{\ast}, \Delta \}
       \exp \left\{ - \frac{g}{\hbar} \int_0^{\hbar\beta } d\tau \int d^3x
            \right.
\]
\[
    \left.  \left| \psi_{\downarrow} \psi_{\uparrow}
     +  \frac{1}{g} \Delta (x,\tau) \right|^2  \right\} \; ,  
\]
into Eq. (20) we have 
\bea
   Z  & = &   \int {\cal D}\{x_v, \psi^{\dag}, \psi, 
        \Delta^{\ast}, \Delta \}
      \exp \left\{  - \frac{1}{\hbar} \int_0^{\hbar\beta } d\tau\int d^3x 
            \right.  \nonumber \\
  & &  \left( \psi^{\dag}_{\uparrow}, 
              \psi_{\downarrow} \right )
        ( \hbar\partial \tau + {\cal H} ) \left( \begin{array}{c} 
                       \psi_{\uparrow}               \\
                       \psi^{\dag}_{\downarrow} \\
                       \end{array}  \right)    \nonumber  \\  
  & &  \left.  - \frac{1}{\hbar g}  \int_0^{\hbar\beta } d\tau \int d^3x  
         |\Delta |^2   \right\} \; . 
\eea
Here the Hamiltonian is defined as
\be
   \ba{l}
   {\cal H}( \Delta, \Delta^{\ast} ) = \left( \begin{array}{cc} 
                     H & \Delta  \\
                    \Delta^{\ast} & - H^{\ast} 
                   \end{array} \right) \; 
   \ea
\ee
with $H =  - (\hbar^{2}/2m ) \nabla^{2} -  \mu_F + V(x) + U_0(x-x_v) $.

Exactly integrating out the electron
fields $\psi_{\sigma}^{\dag}$ and $ \psi_{\sigma}$ first, 
then integrating out 
the auxiliary (pair) fields $\Delta$ under the mean-field approximation, 
one obtains the partition function for the vortex 
   \bea   
   Z = \int {\cal D}\{ x_v \}
       \exp \left\{ - \frac{ S_{eff} }{ \hbar }  \right\} \; 
   \eea
with the effective action
   \bea
   \frac{S_{eff} }{\hbar} = - Tr \ln G^{-1} + 
     \frac{1}{\hbar g}\int_0^{\hbar\beta} d\tau \int d^{3}x|\Delta|^{2} \; ,
   \eea 
where $Tr$ includes internal and  space-time indices,
and the Nambu-Gor'kov Green's function $G$ is defined by 
\be
     ( \hbar \partial_\tau + {\cal H})
      G(x,\tau; x',\tau') = \delta(\tau-\tau') \delta^3(x-x') \, , 
\ee
together with the BCS gap equation, or the self-consistent equation,
\be
  \ba{l}
     \Delta(x,\tau) = \hbar g \;  G_{12}(x,\tau; x,\tau) \; .
   \ea
\ee
In the presence of impurity potentials, the averaging over them 
is implied in Eq. (26), unless explicitly specified.

Since the effective action is a smooth function of vortex
coordinate in the functional space of $\{ x_v(\tau) \}$,
we consider 
that the vortex has made a small move from  its original place  
$x_0$, 
which allows a small parameter expansion in terms of the difference 
between the vortex position $x_v$ and $x_0$. 
We look for the long-time behavior of vortex dynamics under this small
parameter expansion. In the final step, the forces on vortex are 
to be calculated by
varying Lagrangian to this small motion.
The influence of the eliminated degrees of freedom on the vortex dynamics
will then be obtained.
As an example, for the mean-field value of the order parameter, this 
small parameter expansion to the second order is
\bea
   \Delta(x,\tau, x_v) & = &  \left( 1 
         + \delta x_v(\tau) \cdot \nabla_{x_0} 
     + \frac{1}{2}  ( \delta x_v(\tau) \cdot \nabla_{x_0} )^2 \right) \times
              \nonumber  \\
     & &   \Delta_0(x,x_0) \; .
\eea
Here $\delta x_v = x_v - x_0$.
In Eq. (27) we have used the fact that when $x_v = x_0$  the vortex is static.
The effective action for the vortex to the same order is, 
after dropping a constant term,
\bea
     \frac{S_{ eff } }{\hbar }  & = &   \frac{1}{2} Tr (G_0 \Sigma' )^2 
          + \frac{1}{\hbar g} \int_0^{\hbar\beta} d\tau \int d^3x
              \nonumber   \\
   & &   \; 
          \delta x_v\cdot \nabla_{x_0} \Delta^{\ast}_0 \; 
                 \delta x_v\cdot \nabla_{x_0} \Delta_0 \; 
\eea
with
\be
   \ba{l} 
    \Sigma' = \delta x_v\cdot \nabla_{x_0}
              \left( \begin{array}{cc}
               U_0  & \Delta_0 \\
              \Delta^{\ast}_0 & - U_0
               \end{array} \right)
            =  \delta x_v \cdot \nabla_{x_0} {\cal H}_0  \; .
   \ea
\ee
Here the Hamiltonian ${\cal H}_0 = {\cal H}|_{x_v = x_0 }$ for 
the static vortex at $x_0$, 
$G_0$ is the Nambu-Gor'kov Green's function with ${\cal H} $
replaced by  ${\cal H}_0 $,
the gradient $\nabla_{x_0}$ is with respect to $x_0$,
and  $ G^{-1} = G^{-1}_{0} + \Sigma' $.

Now we construct the Nambu-Gor'kov Green's function $G_0$ 
following the usual procedure.\cite{schrieffer}
First, we consider the eigenfunctions of ${\cal H}_0 $. 
The stationary equation, the Bogoliubov-de Gennes equation, is 
\be
   \ba{l}
   {\cal H}_0 \Psi_{\alpha}(x) = E_{\alpha} \Psi_{\alpha}(x) \; ,
   \ea
\ee
with 
\[
   \ba{l}
   \Psi_{\alpha}(x) = \left( \begin{array}{c} u_{\alpha}(x) \\ v_{\alpha}(x) \end{array} 
       \right) \; .
   \ea
\]   
No confusion with the vortex velocity in Sec. II should arise here. 

Given  the eigenfunctions of Eq. (30),
$G_0$ can be expressed as
  \be
   G_0(x,\tau; x',\tau' ) = \sum_{n,\alpha} \frac{-1}{\hbar \beta} 
          \frac{ e^{ - i \omega_n (\tau -\tau') } }{ i\hbar \omega_n - E_{\alpha} }
          \Psi_{\alpha}(x) \Psi_{\alpha}^{\dag}(x') .
  \ee
Here $\omega_n = n \pi / \hbar\beta$, with $n$ odd integers.

Direct substituting of Eq. (31) into Eq. (28) leads to 
\bea
   & & \frac{ S_{eff} }{ \hbar} \nonumber \\
   & & =  \frac{1}{2(\hbar\beta)^2} \int d^3x d^3x' d\tau d\tau' 
   \sum_{n\alpha, n'\alpha'}  
   \frac{e^{-i\omega_n(\tau-\tau')}}{i\hbar\omega_n - E_{\alpha}}
   \frac{e^{-i\omega_{n'}(\tau'-\tau)}}{i\hbar\omega_{n'} - E_{\alpha'}} 
       \times \nonumber \\
 & & {\ } {\ }
   \Psi_{\alpha}^{\dag}(x')\delta x_v(\tau ')\cdot\nabla_{x_0}
    {\cal H}_0(x') \Psi_{\alpha'}(x')    \times \nonumber \\
 & & {\ } {\ }
    \Psi_{\alpha'}^{\dag}(x)\delta x_v(\tau ) \cdot\nabla_{x_0}
   {\cal H}_0(x) \Psi_{\alpha}(x)     \nonumber \\
 & &  { \ } +\frac{1}{\hbar g} \int_0^{\hbar\beta} d\tau \int d^3x 
      \delta x_v(\tau) \cdot \nabla_{x_0} \Delta^{\ast}_0 \; 
                 \delta x_v(\tau) \cdot \nabla_{x_0} \Delta_0 \; .
\eea
Keeping only terms relevant to vortex dynamics and
assuming global rotational symmetry after summing over all the
states,  we have
\bea
   & & \frac{ S_{eff} }{ \hbar } \nonumber \\
   & & = 
   \frac{1}{2(\hbar\beta)^2} \int d^3x d^3x' d\tau d\tau' 
   \sum_{n\alpha,n'\alpha'}\frac{e^{-i\omega_n(\tau-\tau')}}
    {i\hbar\omega_n - E_{\alpha}}
   \frac{e^{-i\omega_{n'}(\tau'-\tau)}}{i\hbar\omega_{n'} - E_{\alpha'}}  
     \times \nonumber \\
  & & {\ }  \left[ \, \Psi_{\alpha}^{\dag}(x')\nabla_0
   {\cal H}_0 \Psi_{\alpha'}(x')\cdot
   \Psi_{\alpha'}^{\dag}(x) \nabla_{x_0}
   {\cal H}_0 \Psi_{\alpha}(x) \; \right.  \times   \nonumber \\
  & & {\ } {\ }  
      \delta x_v(\tau ')\cdot\delta x_v(\tau ) 
    + \left( \Psi_{\alpha}^{\dag}(x')\nabla_{x_0}
     {\cal H}_0 \Psi_{\alpha'}(x') \times  \right. \nonumber \\
  & & {\ } {\ }  \left. \left.  
    \Psi_{\alpha'}^{\dag}(x)\nabla_{x_0}
   {\cal H}_0 \Psi_{\alpha}(x) \right)\cdot \hat{z}  
    \left( \delta x_v(\tau ')\times\delta x_v(\tau ) \right)
            \cdot \hat{z} \, \right]  \, .
\eea
With a rearrangement, finally we arrive at
\bea
    S_{eff}  & = &  \frac{1}{2}  \int_0^{\hbar\beta} d\tau
    \left\{  \int_0^{\hbar\beta} d\tau' 
      F_{\parallel} ( \tau-\tau')|\delta x_v(\tau)- \delta x_v(\tau') |^2  
      \right. \nonumber \\ 
   & &
      \left.      -   \int_0^{\hbar\beta} d\tau' 
            F_{\perp} ( \tau-\tau') 
           (\delta x_v(\tau) \times \delta x_v(\tau') )\cdot \hat{z} 
                  \right\}  \; ,
\eea
with
\bea
 & & F_{\parallel} ( \tau-\tau') \nonumber  \\
  & & = - \frac{1}{2\hbar\beta^2}
  \int d^3x d^3x' 
  \sum_{n\alpha, n'\alpha'}\frac{e^{-i\omega_n(\tau-\tau')}}
    {i\hbar\omega_n - E_{\alpha}}
  \frac{e^{-i\omega_{n'}(\tau'-\tau)}}{i\hbar\omega_{n'} - E_{\alpha'}} 
     \times \nonumber \\
  & & {\ } \Psi_{\alpha}^{\dag}(x')\nabla_{x_0}
  {\cal H}_0 \Psi_{\alpha'}(x')\cdot
  \Psi_{\alpha'}^{\dag}(x)\nabla_{x_0}
  {\cal H}_0 \Psi_{\alpha}(x) \; ,
\eea
and 
\bea
 & &  F_{\perp} ( \tau-\tau') \nonumber \\
 & &  =  \frac{1}{\hbar\beta^2} \int d^3x d^3x'  
 \sum_{n\alpha, n'\alpha'}\frac{e^{-i\omega_n(\tau-\tau')}}
                               {i\hbar\omega_n - E_{\alpha}}
 \frac{e^{-i\omega_{n'}(\tau'-\tau)}}{i\hbar\omega_{n'} - E_{\alpha'}}
    \times \nonumber \\
& & {\ } \left(\Psi_{\alpha}^{\dag}(x')\nabla_{x_0}
 {\cal H}_0 \Psi_{\alpha'}(x')\times
 \Psi_{\alpha'}^{\dag}(x)\nabla_{x_0}
 {\cal H}_0 \Psi_{\alpha}(x) \right) \cdot \hat{z}\, .
\eea
Equation (34) has the form of influence functional in quantum dissipative 
dynamics.\cite{leggett}
Please note that $ \int_{0}^{\hbar\beta} F_{\parallel}(\tau-\tau')d\tau = 0$.
Therefore, there is no so-called `counter-term' in Eq. (33) 
as discussed in Ref.\onlinecite{leggett}.
Here we have generalized the influence functional to 
include the transverse force as a response from the environment.

Before proceeding to evaluate these correlations, 
we discuss some properties of the wave functions
of the Bogoliubov-de Gennes equation, which will be  used later.
First, because ${\cal H}_0$ is Hermitian, all its eigenstates form a  
complete and orthonormal set, that is, 
\[ 
      \int d^3 x \Psi^{\dag}_{\alpha}(x)\Psi_{\alpha'}(x)  
    = \delta_{\alpha,\alpha'} \; 
\]
and 
\[
  \sum_{\alpha} \Psi_{\alpha}(x)\Psi_{\alpha}^{\dag}(x')           
     =  {\bf 1}  \; . 
\]
Here  $\Psi^{\dag}(x) = ( u^{\ast}(x), v^{\ast}(x) ) $ and the wave function
$\Psi(x)$
is normalized to 1 over a cylinder of radius $R$ and length $L$, the box
normalization.  In the thermodynamic limit, $R = \infty$, 
one may consider the scattering states.
In this case the Dirac $\delta$ function normalization for extended
states should be the better choice.
Furthermore, Eq. (30) has the property that if 
\[
    {\cal H}_0 \Psi(x) = E \; \Psi(x)    \; ,  \;    
   \overline{\Psi}(x) = \left( \begin{array}{c} 
                                v^{\ast}(x) \\ 
                      - u^{\ast}(x)\end{array} \right) \; ,
\]
then
\be
   {\cal H}_0 \overline{\Psi}(x) = - E \; \overline{\Psi}(x) \; .
\ee 
There is no specific assumption about the Hamiltonian $H (H^{\ast}) $ 
in Eq. (22) for this identity.
There is another important property implied by Eq. (30).
Since both the Hamiltonian ${\cal H}_0$ and its eigenfunctions 
are the function of the vortex coordinate at $x_0$,
taking the derivative with respect to $x_0$ at both sides of Eq. (30), 
we have
\[ 
  (\nabla_{x_0}{\cal H} ) |\Psi_{\alpha'} \rangle  
   + {\cal H} |\nabla_{x_0} \Psi_{\alpha'} \rangle 
   = E_{\alpha'}   |\nabla_{x_0}\Psi_{\alpha'}  \rangle \; .
\]
Multiplying both sides of this equation by $ \langle \Psi_{\alpha} |$, and
using the relation that 
$ \langle \Psi_{\alpha} |{\cal H} =  E_{\alpha} \langle \Psi_{\alpha} |$, 
the Hermitian conjugation of Eq. (30), for $\alpha \neq \alpha'$  we have  
\bea
  & & \int d^3x \Psi_{\alpha}^{\dag}(x) 
         (\nabla_{x_0} {\cal H}_0)\Psi_{\alpha'}(x)  \nonumber \\
  & &  = (E_{\alpha'} - E_{\alpha})  \int d^3x \Psi_{\alpha}^{\dag}(x)  
         \nabla_{x_0}\Psi_{\alpha'}(x)  \; ,
\eea
with
\[
  \ba{l}
  (\nabla_{x_0} {\cal H}_0) \equiv
    \nabla_{x_0} \left( \begin{array}{cc} 
                    U_0   &  \Delta  \\
                    \Delta^{\ast} & - U_0
                   \end{array} \right) \; .
   \ea
\]
Here we have used $ \nabla_{x_0} E_{\alpha'} = 0 $ to get Eq. (38),
under the assumption that the system is homogeneous.
Hence there is no the vortex velocity independent potential for 
the vortex arising from Eq. (38),
that is, no Born-Oppenheimer type potential, in accordance with
the present purpose of looking for the effects which are first order 
in vortex velocity.
Starting from the Hermitian conjugate of Eq. (30), 
taking the derivative with respect to the vortex coordinate we have
\[
  \langle \nabla_{x_0} \Psi_{\alpha} | {\cal H} 
  + \langle \Psi_{\alpha} | ( \nabla_{x_0} {\cal H} )
  =  E_{\alpha} \langle  \nabla_{x_0} \Psi_{\alpha} | \; .
\]
Then multiplying this equation by $|\Psi_{\alpha'} \rangle$ we have
\bea
    & & \int d^3x \Psi_{\alpha}^{\dag}(x)  
        (\nabla_{x_0} {\cal H}_0)\Psi_{\alpha'}(x) \nonumber \\
    & &  =  - (E_{\alpha'} - E_{\alpha})  \int d^3x  
        \nabla_{x_0} \Psi_{\alpha}^{\dag}(x) \Psi_{\alpha'}(x)  \; .
\eea
We note that both Eqs. (38) and (39) are exact, 
following from the general property of Eq. (30). 
They relate the transition elements of the Hamiltonian after
the differentiation with respect to a parameter to the 
connections between wave functions.  
Though the wave functions have to be determined as an eigenvalue problem,
the usefulness of Eqs. (38) and (39) is that it allows one to concentrate
on wave functions instead of the original Hamiltonian,
which is particularly convenient in the discussion of certain topological
properties described better by wave functions, 
such as a vortex in a BCS superfluid here.
In the rest of the paper, we will take the trapping potential to be zero,
$ U_0 \rightarrow 0$ unless specified,
and determine the vortex position self-consistently through the gap 
equation, Eq. (26).

For the convenience of calculation, sometimes we wish to
use $ \nabla \Psi_{\alpha}(x)$ instead of $\nabla_{x_0} \Psi_{\alpha}(x) $ 
in the expression. It can be done in the following way. 
We split  the gap function, or the order parameter $ \Delta $ into
\[  
  \Delta  = \bar{\Delta}(x-x_0)  + \Delta'(x,x_0)  \; ,
\]
where $\bar{\Delta}$ is a smooth part of the self-consistent potential,  
$\Delta'$ is the fluctuating part
for a given impurity configuration. 
The impurity average gives  $<\Delta'>= 0$.
In the presence of impurity potentials the gap function 
$\bar {\Delta}$ may differ from the one in the clean limit.
The Hamiltonian becomes ${\cal H}_0= \bar{{\cal H}_0} + \delta{\cal H}$, 
with 
\be
   \bar{{\cal H}_0} = \left(\begin{array}{cc} 
                     H_0 & \bar\Delta  \\
                    \bar\Delta^{\ast} & - H_0^{\ast} 
                   \end{array} \right) \; ,
\ee
where $H_0 =  - (\hbar^{2}/2m ) \nabla^{2} -  \mu_F $, and
\be
   \delta{\cal H} = \left( \begin{array}{cc} 
                     V(x) & \Delta'  \\
                    \Delta'^{\ast}  & - V(x) 
                   \end{array} \right) \; .
\ee
Using 
\bea 
 & & \int d^3x \Psi_{\alpha}^{\dag}(x)
        \nabla ( {\cal H}_0 \Psi_{\alpha'}(x) )  \nonumber \\
 & & = \int d^3x \left[ 
    \Psi_{\alpha}^{\dag}(x) ( \nabla {\cal H}_0 ) \Psi_{\alpha'}(x) 
    + \Psi_{\alpha}^{\dag}(x)   {\cal H}_0  
       \nabla  \Psi_{\alpha'}(x)\right] \, , \nonumber 
\eea 
and  $( \nabla + \nabla_{x_0} ) \bar{\cal H}_0 = 0$,  
and defining
\[
 \nabla {\cal H}' \equiv  ( \nabla + \nabla_{x_0} ) \delta {\cal H} \; ,
\]   
we have the desired relations
\bea
    & & (E_{\alpha'} - E_{\alpha})  \int d^3x \Psi_{\alpha}^{\dag}(x)  
      \nabla_{x_0} \Psi_{\alpha'}(x) \nonumber \\
  & & = - (E_{\alpha'} - E_{\alpha})  \int d^3x \Psi_{\alpha}^{\dag}(x)  
      \nabla \Psi_{\alpha'}(x) \nonumber \\
  & & {\ } + \int d^3x \Psi_{\alpha}^{\dag}(x)  
      \nabla   {\cal H}'\Psi_{\alpha'}(x) \, 
\eea
and 
\bea
 & & (E_{\alpha'} - E_{\alpha})  \int d^3x   
   \nabla_{x_0} \Psi_{\alpha}^{\dag}(x)   \Psi_{\alpha'}(x) \nonumber \\
 & & = - (E_{\alpha'} - E_{\alpha})  \int d^3x \nabla \Psi_{\alpha}^{\dag}(x)  
       \Psi_{\alpha'}(x) \nonumber \\
 & & {\ } -  \int d^3x \Psi_{\alpha}^{\dag}(x)  
    \nabla   {\cal H}'  \Psi_{\alpha'}(x) \, .
\eea
The last part is obtained from Eq. (42) by a partial integration, which
can be carried through because the wave function is normalizable,
either by the box normalization or by the Dirac delta function.

\subsection{ Longitudinal Correlation}

We now discuss the general properties of
the longitudinal correlation function, Eq. (35).
We find
\bea
 & & \sum_{n,n'} 
 \frac{e^{-i\omega_n(\tau-\tau')}}{i\hbar\omega_n - E_{\alpha}}
 \frac{e^{-i\omega_{n'}(\tau'-\tau)}}{i\hbar\omega_{n'} - E_{\alpha'}}  
 \nonumber \\
 & & = \sum_{n,n'} \frac{e^{-i(\omega_n -\omega_{n'})(\tau-\tau')}}
 { i(\hbar\omega_n -\hbar\omega_{n'}) - (E_{\alpha}- E_{\alpha'})}  \times
        \nonumber \\
 & & {\ } \left( \frac{1}{i\hbar\omega_{n'}- E_{\alpha'}} -  
   \frac{1}{i\hbar\omega_{n}- E_{\alpha}}\right)
             \nonumber \\
 & & =  \sum_{n-n'}\beta (f_{\alpha'}-f_{\alpha})
  \frac{e^{-i(\omega_n -\omega_{n'})(\tau-\tau')}}
  { i(\hbar\omega_n -\hbar\omega_{n'}) - (E_{\alpha}- E_{\alpha'})}   \, ,
\eea
after using 
\bea
     \sum_n \frac{ e^{ -i \omega_n \delta}  }
                 { i\hbar \omega_n - E_{\alpha}    } = 
     \left\{
        \ba{cc}
       \beta \; f_{\alpha} \; ,      & \delta = 0^-  \nonumber \\
       - \beta \; (1 - f_{\alpha} )\; , & \delta = 0^+  \nonumber \\    
         \ea
   \right.  \; ,
\eea
with the Fermi distribution function 
$f_{\alpha} = 1/(1 + e^{\beta E_{\alpha}} )$.
To complete the calculation,  we also need 
\bea
& & \sum_{n-n'}
 \frac{ \cos[ (\omega_n -\omega_{n'})(\tau-\tau') ]}
 { i(\hbar\omega_n -\hbar\omega_{n'}) - (E_{\alpha}- E_{\alpha'})}  
    \nonumber \\  
 & & = -\frac{\beta}{2}  
 \frac{ \cosh\left[ \frac{(E_{\alpha} - E_{\alpha'}) }{\hbar} 
   \left(\frac{\hbar\beta}{2} - |\tau-\tau' | \right) \right] }
     {\sinh\left[ (E_{\alpha} - E_{\alpha'}) \frac{\beta}{2} \right] } 
  + :\delta(\tau - \tau): \, .
\eea
Here $ :\delta(\tau): $ is a periodic delta function with 
period $\hbar\beta$.
The term with 
$ \sum_{n-n'}  { \sin[ (\omega_n -\omega_{n'})(\tau-\tau') ]} /
  { [ i(\hbar\omega_n -\hbar\omega_{n'}) - (E_{\alpha}- E_{\alpha'}) ] } $
is zero inside the double imaginary time integration in Eq. (34), 
because the integrand is an odd function of $ \tau-\tau'$.
Dropping the periodic $\delta$ function, whose contribution
is zero in Eq. (34), we are ready then to write down 
the longitudinal correlation function as
\be  
  F_{\parallel}(\tau) = \frac{1}{\pi} \int^{\infty}_{0} d\omega J(\omega)
    \frac{\cosh\left[\omega\left(\frac{\hbar\beta }{2} - |\tau|\right)\right]}
         {\sinh\left[\omega\frac{\hbar \beta}{2}\right]} \, , 
\ee
with  the spectral function
\bea
   J(\omega)& = & \frac{\pi }{4} \sum_{\alpha,\alpha'}
    \delta(\hbar\omega - | E_{\alpha} - E_{\alpha'} |)
     | f_{\alpha'} -  f_{\alpha} |       \times   \nonumber \\
    &  & \left| \int d^3x \Psi_{\alpha}^{\dag}(x)
            \nabla_{x_0}{\cal H}_0 \Psi_{\alpha'}(x) \right|^2 \; .
\eea
It is interesting to point out that 
in terms of the spectral function the longitudinal 
correlation function, Eq. (46), is in exactly the same form 
of the influence functional in quantum dissipative dynamics.\cite{leggett}
The apparent difference is that the spectral function 
in Ref.\onlinecite{leggett} has been obtained by integrating out 
a set of independent harmonic oscillators (bosons), while here
it has come from the elimination of independent fermionic modes
determined by the Bogoliubov-de Gennes equation.

It is important to remember that in order to have a 
smooth spectral function $J(\omega)$,
the thermodynamic limit must be taken first before the implementation
of the $\delta$ function in Eq. (47).\cite{za}
Otherwise the spectral function consists of sum of 
discrete delta functions, and there would be no dissipation.
This limit procedure is 
in accordance with the requirement in nonequilibrium statistical mechanics: 
The thermodynamic limit must be taken first to have well-defined 
low-lying modes in the zero-frequency limit.
After this consideration of the thermodynamic limit, 
in the low-frequency limit the spectral function 
may have the following generic form:
\be
  J(\omega) = \eta \; \omega^s   \; , \; \omega \rightarrow 0^{+} \; ,
\ee
with $ s > 1$ being the super-ohmic case,  $ 1> s > 0$ being 
the sub-ohmic case, and $s=1$ being the Ohmic case,
following from the influence functional formulation 
of quantum dissipative dynamics.\cite{leggett}
For the physically important Ohmic case, 
the longitudinal force, friction, is given by 
$- \eta  {\bf v}_V$, 
and from Eq. (47) we have the frictional coefficient
\bea
 \eta & = & \frac{\pi }{4}
    \sum_{ \alpha' \ne \alpha }   \hbar  
     \frac{f_{\alpha} - f_{\alpha'} }{ E_{\alpha'}- E_{\alpha}  }
  \delta( 0^{+} - | E_{\alpha}- E_{\alpha'}| )  \times    \nonumber \\
  & &  |\langle\Psi_{\alpha}|\nabla_{x_0} {\cal H}_0 |\Psi_{\alpha' } 
   \rangle |^2  \, .
\eea
This equation is the familiar Fermi Golden rule for dissipation.
The matrix elements of $\nabla_{x_0} {\cal H}_0 $ are well behaved.
If we use  Eqs. (38) and (39) to re-express the frictional coefficient $\eta$
in terms of the overlap integral between the wave functions
 $  | \nabla_{x_0} \Psi_{\alpha} \rangle $ and $ |\Psi_{\alpha' } \rangle $,
we turn it into the form of ratio 
$0/0$ when $E_{\alpha} - E_{\alpha'} \rightarrow 0$.
Then attention should be paid to the divergence of 
the overlap integral when $ | E_{\alpha}- E_{\alpha'}| \rightarrow 0^{+}$.
This limiting behavior has been discussed in Ref.\ \onlinecite{suhl}, and
we refer the reader to the Appendix for a detailed discussion.
 Equation (49) clearly shows that 
the coefficient of friction $\eta$ is  determined by 
low-energy excitations such as phonons, extended quasiparticles,
and bounded core quasiparticles when their energy spectrum is smeared out by
impurities. 
The equivalence of Eq. (49) in the context of vortex dynamics  
to a more conventional partial wave phase-shift analysis
has been discussed in Ref.\onlinecite{tt} for a few well-defined situations.
A more formal discussion can be found in Ref. \onlinecite{suhl}.
It may also be instructive to mention here that 
the friction experienced by a moving object in a normal Fermi liquid 
has been analyzed in the influence functional approach.\cite{chen}
Those considerations
suggests that nonzero extended states friction contributions exist, as 
will be borne out in detail in the next section.
Finally, 
it should also be pointed out that Eq. (49) is a special case of Eq. (47). 
It will not pick up any super-Ohmic contributions, 
and will give infinity for any sub-Ohmic contributions.
If such cases occur, we need to return to the general 
expressions, Eqs. (46) and (47).

To close this subsection, there are two general remarks in order.
First, the present result of expressing the frictional coefficient
in terms of low-lying excitations is in accordance with Landau's quasiparticle
picture: In the zero-frequency limit the lifetime of those excitations 
approaches infinity. Those excitations give an exact
description of the dynamics of the whole system in this limit.
Secondly, to relate to the discussion 
in Sec. II.B, the spectral function given by Eq. (47) 
completely determines the spectral representation of 
the second kind of fluctuation-dissipation theorems. 
Equation (47) is indeed a quantitative description of dissipation.

\subsection{ Transverse Correlation }

To obtain physically more transparent expressions for 
the transverse correlation function in Eq. (36),
we use Eqs. (38) and (39) to rewrite it as
\bea
 & &  F_{\perp} ( \tau-\tau')   \nonumber  \\
 & & =  \frac{1}{\hbar\beta^2} \int d^3x d^3x'  
   \sum_{n\alpha,n'\alpha'}
   \frac{e^{-i\omega_n(\tau-\tau')}}{i\hbar\omega_n - E_{\alpha}}
   \frac{e^{-i\omega_{n'}(\tau'-\tau)}}
   {i\hbar\omega_{n'} - E_{\alpha'}} \times  \nonumber \\
 & & {\ } {\ } (E_{\alpha}-E_{\alpha'})^2 \times
          \nonumber \\
 & & {\ } \left[ \left( \Psi_{\alpha}^{\dag}(x')\nabla_{x_0} 
           \Psi_{\alpha'}(x') \right) \times
   \left(\nabla_{x_0}\Psi_{\alpha'}^{\dag}(x) \; 
       \Psi_{\alpha}(x) \right)\right]
   \cdot \hat{z} \; .
\eea
According to Eq. (44) and (45), 
\bea
 & & \sum_{n,n'}
 \frac{e^{-i(\omega_n - \omega_{n'})(\tau - \tau')}}{
 (i\hbar\omega_n  - E_{\alpha})(i\hbar\omega_{n'} - E_{\alpha'})} \nonumber \\
 & & = \beta\sum_n \frac{e^{-i\tilde{\omega}_n(\tau - \tau')}}
 { i\hbar\tilde{\omega}_n - ( E_{\alpha} -  E_{\alpha'})}
 (f_{\alpha'} - f_{\alpha})  \nonumber \\
 & & = \frac{\beta^2}{2} 
   \left[ -1 
   + \frac{ \hbar}{ E_{\alpha} -  E_{\alpha'} } \partial_{\tau - \tau'} \right]
        \nonumber \\ 
 & & {\ }  \frac{\cosh\left[ \frac{ E_{\alpha} -  E_{\alpha'} }{\hbar}
              \left(\frac{\hbar\beta}{2} - |\tau - \tau'|\right)   \right] }
        {\sinh\left[ \frac{ (E_{\alpha} -  E_{\alpha'})\beta }{2} \right]  }
      (f_{\alpha'} - f_{\alpha})  \; . \nonumber 
\eea
Because of the symmetry with the interchange of $\alpha$ and $\alpha'$,
the $ - 1$ term in the above square brackets does not contribute 
to the transverse correlation function, and we have
\bea
   & & F_{\perp} ( \tau-\tau')     \nonumber  \\
 & & = \frac{1}{2} \int d^3x d^3x'  \sum_{\alpha, \alpha'}
     \partial_{\tau - \tau'} 
   \frac{\cosh\left[ \frac{ E_{\alpha} -  E_{\alpha'} }{\hbar}
              \left(\frac{\hbar\beta}{2} - |\tau - \tau'|\right)   \right] }
        {\sinh\left[ \frac{ (E_{\alpha} -  E_{\alpha'})\beta }{2} \right]  }
      \nonumber \\
 & & (E_{\alpha} - E_{\alpha'}) ( f_{\alpha'} - f_{\alpha} ) \times
          \nonumber \\
 & & \left[ \left( \Psi_{\alpha}^{\dag}(x')\nabla_{x_0} 
           \Psi_{\alpha'}(x') \right) \times
   \left(\nabla_{x_0}\Psi_{\alpha'}^{\dag}(x) \; 
       \Psi_{\alpha}(x) \right)\right]
   \cdot \hat{z} \; . \nonumber 
\eea
The corresponding term in the effective action, Eq. (34), is then
\bea 
   & & - \frac{1}{2} \int_{0}^{\hbar\beta} d\tau \int_{0}^{\hbar\beta} d\tau'
      F_{\perp}(\tau-\tau') ( \delta x_{v}(\tau) \times \delta x_v(\tau'))
             \cdot \hat{z}       \nonumber \\
   & & = - \frac{1}{4}\int_{0}^{\hbar\beta} d\tau \int_{0}^{\hbar\beta} d\tau'
                 \int d^3x d^3x'  \sum_{\alpha, \alpha'}
         \nonumber \\
   & & {\ } {\ }
    \partial_{\tau - \tau'} 
   \frac{\cosh\left[ \frac{ E_{\alpha} -  E_{\alpha'} }{\hbar}
              \left(\frac{\hbar\beta}{2} - |\tau - \tau'|\right)   \right] }
        {\sinh\left[ \frac{ (E_{\alpha} -  E_{\alpha'})\beta }{2} \right]  }
      \nonumber \\
 & & {\ } {\ } (E_{\alpha}-E_{\alpha'}) ( f_{\alpha'} - f_{\alpha} ) \times
          \nonumber \\
 & & {\ } \left[ \left( \Psi_{\alpha}^{\dag}(x')\nabla_{x_0} 
           \Psi_{\alpha'}(x') \right) \times
   \left(\nabla_{x_0}\Psi_{\alpha'}^{\dag}(x) \; 
       \Psi_{\alpha}(x) \right)\right]
   \cdot \hat{z} \times 
         \nonumber \\
  & & {\ } {\ }
   ( \delta x_{v}(\tau) \times \delta x_v(\tau'))
             \cdot \hat{z}  \; . \nonumber   \\
 & & = - \frac{1}{4} \int_{0}^{\hbar\beta} d\tau 
                     \int_{-\infty}^{\infty} d\tau'
                 \int d^3x d^3x'  \sum_{\alpha, \alpha'}
      \nonumber \\
  & & {\ } {\ } \partial_{\tau - \tau'} 
    \exp\left\{ - \frac{ |E_{\alpha} -  E_{\alpha'}| }{\hbar}
              |\tau - \tau'|  \right\} 
        sgn( E_{\alpha} -  E_{\alpha'} )
      \nonumber \\
 & & {\ } {\ } (E_{\alpha}-E_{\alpha'}) ( f_{\alpha'} - f_{\alpha} ) \times
          \nonumber \\
 & & {\ } \left[ \left( \Psi_{\alpha}^{\dag}(x')\nabla_{x_0} 
           \Psi_{\alpha'}(x') \right) \times
   \left(\nabla_{x_0}\Psi_{\alpha'}^{\dag}(x) \; 
       \Psi_{\alpha}(x) \right)\right]
   \cdot \hat{z} \times \nonumber \\
  & & {\ } {\ } ( \delta x_{v}(\tau) \times \delta x_v(\tau'))
             \cdot \hat{z}  \; . 
\eea
In the last equality we have used the periodicity of the function
$\delta x_v(\tau) = \delta x_v(\hbar\beta + \tau) $ to 
turn the hyperbolic function into the exponential function.
Now we look for the slow motion expansion to the leading order
in velocity:
 $\delta x_v(\tau') = \delta x_v(\tau) + \delta \dot{x}_v(\tau) ( \tau'-\tau)$.
Substituting this expansion into Eq. (51), after the integration over
$ (\tau'-\tau) $ we have 
\bea
 & &  - \frac{1}{2} \int_{0}^{\hbar\beta} d\tau 
                    \int_{ 0 }^{ \hbar\beta } d\tau'
      F_{\perp}(\tau-\tau') ( \delta x_{v}(\tau) \times \delta x_v(\tau'))
             \cdot \hat{z}   \nonumber \\
 & & =  i \int_{0}^{\hbar\beta} d\tau \, B \, [\delta x_v(\tau) 
          \times \delta \dot{x}_v(\tau) ]\cdot \hat{z}   \; ,
   \nonumber 
\eea
with the quantity $B$ which determines the transverse force defined as 
\bea
   & B = &  i \frac{\hbar}{2} \sum_{\alpha,\alpha'} 
            ( f_{\alpha'} - f_{\alpha} ) 
       \int d^3x \int d^3x'    \nonumber \\
   & &  \hat{z}\cdot \left(  \Psi_{\alpha}^{\dag}(x') 
             \nabla_{x_0}\Psi_{\alpha'}(x')\times
       \nabla_{x_0}\Psi_{\alpha'}^{\dag}(x)  \Psi_{\alpha}(x) \right) \; .
\eea

We demonstrate next that the contribution to the transverse correlation
can be evaluated by counting extended states contributions.
First, we regroup terms in Eq. (52): 
\bea
 B & = &  i \frac{\hbar}{2} \hat{z} \cdot tr \sum_{\alpha,\alpha'}
   ( - ) 2 \sum_n \frac{1}{\beta} 
            \frac{ e^{-i\omega_n \delta } }{ i\hbar\omega_n - E_{\alpha} } 
        \nonumber \\
  & & \int d^3x \int d^3x' \Psi_{\alpha}(x) \Psi_{\alpha}^{\dag}(x') 
     \nabla_{x_0}\Psi_{\alpha'}(x')\times
  \nabla_{x_0}\Psi_{\alpha'}^{\dag}(x)  \nonumber \\
  &  = &  i \frac{\hbar}{2}  \hat{z} \cdot  tr \sum_{\alpha}
  ( - ) 2 \sum_n \frac{1}{\beta} 
            \frac{ e^{-i\omega_n \delta } }{ i\hbar\omega_n - E_{\alpha} } 
            \times    \nonumber \\ 
 & &   \int d^3x \left( \nabla_{x_0} \Psi_{\alpha}(x)\times
       \nabla_{x_0}\Psi_{\alpha}^{\dag}(x) \right) \, ,   
\eea
because that $\{ \Psi_{\alpha'} \} $ form a completed set. 
Here $tr$ stands for summing over spinor indices.  
The replacement of $f_{\alpha}$ by the summation is to take care of the
delicate equal-time limit in the trace: 
$\delta = 0^{-}$ for spin up and  
$\delta = 0^{+}$ for spin down in Nambu spin space. 
We encounter such a choice of time limit only in the case of
taking the trace of the Nambu-Gor'kov Green's function directly.
This choice will not be there if 
we only need to take trace of higher powers 
of the Nambu-Gor'kov Green's function,
 e.g., $tr G_0^2$.
This implies that the functions containing 
occupation numbers $\{ f_{\alpha} \}$ 
in their differences are well defined.
Therefore, Eq. (52) can be safely used if we directly substitute in
eigenstates of the Bogoliubov-de Gennes equation.
An alternative natural way of deriving Eq. (53) is to leave the
summation over $\omega_n$ in place throughout Eq. (50) to Eq. (53). 

After substituting Eq. (53) into Eq. (52),
 we  will write it explicitly in terms of the 
 eigenstates of Bogoliubov-de Gennes equation. Since
\bea 
& &  tr\sum_{\alpha}
   \sum_n \frac{1}{\beta} 
            \frac{ e^{-i\omega_n \delta } }{ i\hbar\omega_n - E_{\alpha} }    
  \int d^3x \nabla_{x_0}\Psi_{\alpha}(x)\times
  \nabla_{x_0}\Psi_{\alpha}^{\dag}(x)  \nonumber \\
 &  &  = \sum_{\alpha}\int d^3x \left\{ f_{\alpha} \nabla_{x_0}  u_{\alpha}(x) 
       \times \nabla_{x_0} u_{\alpha}^{\ast}(x)   \right.  \nonumber \\
 & & {\ } \left.  - (1-f_{\alpha}) \nabla_{x_0} v_{\alpha}(x) \times 
      \nabla_{x_0} v_{\alpha}^{\ast}(x) \right\} 
              \;  , \nonumber 
\eea
we  obtain
\bea
  B  & = & - i \hbar\hat{z}\cdot \sum_{\alpha}
                                    \int d^3x  
       ( f_{\alpha} \nabla_{x_0}u_{\alpha}^{\ast}(x)\times \nabla_{x_0} 
      u_{\alpha}(x) \nonumber \\   
  &  &  - (1-f_{\alpha}) \nabla_{x_0}v_{\alpha}^{\ast}(x)\times 
              \nabla_{x_0} v_{\alpha}(x) )  \; . 
\eea
After using $\nabla_{x_0} \rightarrow -\nabla$, we evaluate Eq. (54)
with the help of the  current definition,\cite{bardeen}
\[ 
  {\bf j} = - \frac{i\hbar}{2} \sum_{\alpha} 
  \left\{ f_{\alpha} u_{\alpha}^{\ast} \nabla u_{\alpha} 
   + (1-f_{\alpha}) v_{\alpha} \nabla v_{\alpha}^{\ast} \right\} + c.c.  \, .
\] 
Equation (54) becomes
\bea
  B & = & \int dx \; \hat{z}\cdot(\nabla\times{\bf j}) \nonumber \\
    & = & \oint_{|x - x_0| \rightarrow \infty }
            d{\bf l}\cdot{\bf j} \nonumber \\
    & = & 2\pi \hbar\rho_s(T)    \, . 
\eea
Here we have used the fact that the current is zero at the vortex position,
as the explicit calculation in Ref. \onlinecite{bardeen} has shown.
It can be understood as the requirement of quantum mechanics:
At the phase singular point, the amplitude of any wave function carrying
this singular phase must be zero. 
In reaching Eq. (55) we have also made the assumption that
the vortex does not have a normal fluid circulation.
The normal fluid is in equilibrium with the substrate or the walls of 
the container.

>From Eq. (55) one may conclude that when counting the 
contribution from individual states, only extended states
give rise to the contribution to the transverse response, because the loop
of the line integral can be chosen arbitrarily large
to make the core state contributions arbitrarily small.
It corresponds to the fact that only extended states can contribute
to the Berry phase of the vortex.\cite{at}
This result is valid even when the trapping potential $U_0$ is finite, which 
we demonstrate here.
Since Eqs. (38) and (39) are valid in the presence 
of a finite trapping potential,
the transverse correlation function can be expressed by wave functions in
exactly the same form as that of vanishing trapping potential, up to Eq. (54).
The wave functions, particularly those for core states,
may be strongly affected by the trapping potential, 
and may even become ill defined.
An example may be the trapping of a vortex by a physical wire.
Now, one may perform the same calculation of turning the area integration
into line integrations, as done in Eq. (55).
Since the trapping potential will not affect the superfluid number 
density far away from the vortex, and since the circulation current 
is still zero at the vortex position,
one then gets the same result as Eq. (55) 
in the presence of a trapping potential.

The validity of Eq. (55) in the presence of a finite trapping potential 
implies that the transverse force is independent of the trapping potential
$ U_0$ at the vortex center.
In Eq. (29), the main function of  
the trapping potential is to specify the vortex position,
a symmetry breaking in an otherwise homogeneous system.
This is similar to the symmetry breaking by an infinitesimal field
near a continuous phase transition in statistical mechanics. 
Hence, it can be effectively taken to be zero, as we have explicitly done
in the present paper.

Next, we turn to the calculation of the superfluid number density $\rho_s$.
At zero temperature, it is straightforward.
It is equal to the total fluid number density 
$\rho_0 =  \sum_{\alpha, E_{\alpha} > 0 } |v_{\alpha}(|x - x_0| \rightarrow \infty)|^2 $,
the number of Cooper pairs per unit area.
At finite temperatures, there is a reduction of superfluid number density
due to the backflow carried by quasiparticle excitations.
In principle one may directly 
calculate the current density together with 
the gap equation, or self-consistent equation, to find out $\rho_s$.
This would be prohibitively difficult.
Instead, one may proceed in the following manner:
Far away from the vortex core, the current varies slowly.
One may take the current to be locally uniform.
Following the same way as that in superfluid He3 using the backflow 
contribution,\cite{vw} 
the superfluid number density can be found as
\be
   \rho_s(T) = \rho_0 ( 1 - Y_0(T) ) \; ,
\ee
with the Yosida function $Y_0$ defined as
\[
  Y_0(T) = \int^{\infty}_{- \infty} d \epsilon  \; 
        \frac{ e^{ \sqrt{\epsilon^2 + \Delta_\infty^2 }/k_B T } }
      {k_B T \left( e^{ \sqrt{\epsilon^2 
                    + \Delta_\infty^2 }/k_B T } + 1 \right)^2 } \; ,
\] 
which accounts for the quasiparticle excitations contributions.
At the superconducting transition temperature $T_c$, $\Delta_{\infty} = 0$ 
and $Y_0(T_c) = 1$, the superfluid number density is zero as expected.   
This expression is the same as that obtained from the London 
penetration depth for a clean type-II superconductor.\cite{tinkham}

Using Eq. (55), the transverse term in the effective action, Eq. (34), is 
\bea
 & & - \frac{1}{2}\int d\tau d\tau' F_{\perp}(\tau-\tau')
  (\delta x_v(\tau) \times \delta x_v(\tau') )\cdot \hat{z}     \nonumber  \\
 & & =  i \int d\tau d\tau' \, B \, 
  (\delta x_v(\tau) \times \delta \dot{x}_v(\tau) )\cdot \hat{z} \nonumber  \\
 & & = - i 2\pi\hbar \rho_s \int d\tau \; 
     \delta \dot{x}_v(\tau)\cdot{\bf A}_t \, 
\eea
with
\[
{\bf A}_t = \frac{1}{2} (\delta x_v\times\hat{z})  \, ,
\]
which has the same form of the action for a charged particle in a uniform
magnetic field.
The geometric phase or the Berry phase for the vortex moving along a
closed trajectory $\Gamma$ is
\bea 
  \Theta & = & 2\pi \hbar \rho_s \int d\tau\; 
     \delta \dot{x}_v(\tau)\cdot{\bf A}_t   \nonumber \\
    & = & 2\pi \hbar \rho_s \oint_{\gamma} 
          d(\delta {x}_v)\cdot{\bf A}_t \nonumber \\
    & = & - 2\pi \hbar \rho_s S(\Gamma)  \;   \nonumber 
\eea
with $ S(\Gamma) $ being the area enclosed by $\Gamma$.
The total transverse force on a vortex is then
\[
  {\bf F} = -2\pi\hbar \rho_s  \delta \dot{x_v}\times\hat{z}  \, .
\]

In view of the foregoing discussions,
we may rewrite our general formulation, Eq. (34),
in a more suggestive form.
The effective action for the vortex is
\bea
    S_{eff}  & = & \left.  \int_0^{\hbar\beta} d\tau 
    \right\{ -i\; 2\pi\hbar \rho_s \delta \dot{x}_v(\tau)\cdot{\bf A}_t 
          \nonumber \\
    & & \left.  + \frac{1}{2}  \int_0^{\hbar\beta} d\tau' 
      F_{\parallel} ( \tau-\tau')|\delta x_v(\tau)- \delta x_v(\tau') |^2  
      \right\}  \; 
\eea
with ${\bf A}_t = \frac{1}{2} (\delta x_v\times\hat{z}) $.
The rewriting of Eq. (46) for the correlation function is
\[
    F_{\parallel}(\tau) = \frac{1}{\pi} \int^{\infty}_{0} d\omega J(\omega)
     \frac{\cosh\left[\omega\left(\frac{\hbar\beta }{2} - |\tau|\right)\right]}
    {\sinh\left[\omega\frac{\hbar \beta}{2} \right]}     \, , 
\]
and the rewriting of Eq. (47) for the spectral function is 
\bea
   J(\omega) & = & \frac{\pi }{4} \sum_{\alpha,\alpha'}
    \delta(\hbar\omega - | E_{\alpha} - E_{\alpha'} |)
     | f_{\alpha'} -  f_{\alpha} | \times \nonumber \\
    & & \left| \int d^3x \Psi_{\alpha}^{\dag}(x)
            \nabla_{x_0}{\cal H}_0 \Psi_{\alpha'}(x) \right|^2 \; . \nonumber
\eea
The thermodynamic limit must be taken first to have a smooth spectral 
function, which is crucial for obtaining a finite vortex friction.

\section{ Vortex Dynamics in Clean Limit }

The example on extremely clean limit of fermionic superfluids 
is the superfluid He 3:
The impurity concentration can be made to be smaller than 1 in $10^{12}$.
For superconductors, the impurity effect can, in principle, be made
arbitrarily small, but no clear experimental realization has been 
reported yet. 
In view of this experimental situation, the discussions 
in this section are more relevant to He 3.
However, from the methodological point of view,
it is instructive to see how the formulation developed in Sec. III
works for such a clean situation.

\subsection{ Extended States Contribution to Vortex Friction }

In this subsection we first calculate the extended state, quasiparticle
and hole excitations, contributions to the vortex friction
to illustrate the usefulness of the present longitudinal response formula.
The formula, Eq. (47) or (49), is formally exact.
However, for a given problem it is difficult to obtain an 
exact detailed expression for friction, 
except in some rare cases.\cite{bonig,chen,tt} 
Hence a WKB-type approximation will be used below. 
The responses of fermions, or electrons, governing by 
Hamiltonian dynamics generates a finite friction for the vortex.

At finite temperatures the extended states above (below) the Fermi level
(the quasiparticles (holes)) are partially occupied. 
The vortex motion causes transitions between these states, 
which gives rise to vortex friction.
The transitions between different single quasiparticle levels
$ \langle\Psi_\alpha|\nabla_{x_0} {\cal H}_0 |\Psi_{\alpha'} \rangle$
are considered here since they dominate the low-energy process.
The quasiparticles
are described by the eigenstates, $u_\alpha $ and $v_\alpha $,
of the Bogoliubov-de Gennes equation. Their behavior in the 
presence of a vortex has been well studied in Ref.\onlinecite{bardeen}. 
We may take
\setcounter{equation}{58}
\be
   \Psi_\alpha   = 
   \left( \begin{array}{c} u_\alpha (x) \\ v_\alpha (x) \end{array} 
       \right) = \frac{ e^{ik_z z} }{\sqrt{L} }
                 \frac{e^{i\mu\theta + i\sigma_z\theta /2} }{\sqrt{2\pi}}
                 \hat{f}(r) \; 
\ee
with ${\bf r}$ measured from the vortex position, $\theta$ is the azimuthal
angle around the vortex, $L$ is the thickness of the superconductor 
film (the length of the vortex line), and $\xi_0$ is the coherence length.
In order to obtain a concrete form for the transition elements,
we use a WKB-type solution for $\hat{f}(r)$
\be
   \hat{f}(r)  = \frac{1}{\sqrt{2} }
   \left( \begin{array}{c} 
   \left[ 1 \pm 
         \frac{ \sqrt{E^2 - |\Delta (r) |^2} }{E} \right]^{\frac{1}{2} } \\ { }
   \left[ 1 \mp 
         \frac{ \sqrt{E^2 - |\Delta (r) |^2} }{E} \right]^{\frac{1}{2} } 
         \end{array} 
       \right) J_{ \mu \pm \frac{1}{2} } ( k_{\pm}(E) r )    \; .
\ee
Here  $k_{\pm}(E) = \sqrt{ k_{\rho}^2 \pm 2m 
      \sqrt{ E^2- |\Delta (r)|^2} /\hbar^2   } $ with  
$k_{\rho}^2 = k_f^2 - k_z^2$.
The negative energy wave functions determined by
the Bogoliubov-de Gennes equation
may be constructed according to Eq. (37) from the positive energy ones.
We will use the approximation that $k_{\rho} \approx k_F$
for the prefactor by assuming that the significant contributions 
come from the region near the Fermi surface.
This WKB-type solution 
is valid when $r$ is outside the classical turning point
$r_t = |\mu|/k_{\rho}$. 
Here $r_t $ is the impact parameter.
A WKB-type solution also exists inside
the turning point. However, because it approaches zero as $ 
(r k_{\rho})^{|\mu|} /|\mu| ! $, the contribution to the transition elements
from this region is small,  and will be set to zero.  
The transition elements are then given by
\begin{eqnarray}
  & & |\langle\Psi_\alpha|\nabla_{x_0} {\cal H}_0 |\Psi_{\alpha' } \rangle|^2
   \nonumber \\
  & & =  \left|\int d^3x ( u_{\alpha'}^{\ast}(x) 
     ( \nabla_{x_0} \Delta )  v_\alpha(x) + 
         v_{\alpha'}^{\ast}(x) 
  (\nabla_{x_0}\Delta^{\ast} ) u_\alpha(x) ) \right|^2  
   \nonumber \\
  & & =   \left\{ \begin{array}{lc} 
      \frac{\Delta_\infty^2}{2\pi^2 \; k_F^2} \; 
         \delta_{k_{z},k_{z}'} \delta_{\mu' ,\mu \pm 1}
   \, , \; & |\mu| \leq \xi_0 k_{\rho} \\
   0 \, , & |\mu|>\xi_0 k_{\rho} \,  \end{array} \right.  \; .
\end{eqnarray}
Here $\Delta_\infty$ is the value of $|\Delta (r)|$ 
far away from the vortex core.
Physically, it means that if the classical quasiparticle trajectory 
is far away from
the vortex core, it will not contribute to the vortex friction.
The summation over states in Eq. (47) or Eq. (49) is replaced by 
\be
  \sum_{{\alpha'}\ne\alpha}=
  \sum_{\mu, \mu', k_{z}, k_{z}'} 
  \int {dE} {dE'} 
    \frac{E}{ \sqrt{E^2 -\Delta_\infty^2} }
   \frac{E'}{\sqrt{E'^2-\Delta_\infty^2} }
  \left(\frac{2m}{\hbar^2 }\right)^2  \; ,
\ee
after considering the density of states. 

Substituting Eqs. (61) and (62) into Eq. (49), 
and using the quasiparticle distribution 
function $f_\alpha  = 1/( e^{\beta E_\alpha} + 1 )$, 
 the coefficient of friction is given by
\be
  \eta =  \frac{Lm^2\xi_0 \Delta_\infty^2\beta}{4\pi^2 \hbar^3} 
   \int_{\Delta_\infty}^{\infty} dE 
   \frac{E^2}{E^2-\Delta_\infty^2  }
     \frac{1}{ \cosh^2{(\beta E/2)}} \;  .
\ee
The integral in Eq. (63) diverges logarithmically. 
It implies that the spectral function
corresponding to the vortex-quasiparticle coupling is not
strictly Ohmic but has an extra frequency factor proportional to
$\ln(\Delta_{\infty}/\hbar \omega )$.
When  $\hbar\omega$ is not very small comparing to $\Delta_\infty$, 
which may be realized   when  close to T$_c$,
we can ignore the logarithmic divergence 
in Eq. (63) by using the density of states
for normal electrons to obtain a finite friction,
i.e., replacing  $E^2/(E^2-\Delta_\infty^2)$ with 1 in Eq. (63).
Close to  T$_c$, the vortex friction approaches zero the same way as 
$\Delta_\infty^2$, which is proportional to the superfluid number 
density $\rho_s$. When $ - \ln(\hbar\omega/\Delta_\infty)$ is large, we
need to use Eq. (47) instead. 
Straightforward evaluation shows that in such a case
\be
   \eta = \frac{Lm^2\xi_0\Delta_\infty^3\beta}{ 8\pi^2\hbar^3 }
   \frac{1}{\cosh^2{(\beta \Delta_\infty /2)}}
   \ln (\Delta_\infty/ \hbar\omega_c ) \, .
\ee
Here $\omega_c$ is the low-frequency cutoff.
It is determined by the size of the system for a single vortex, 
and by the intervortex distance for a vortex array.

We discuss briefly here the connection of our results to previous ones. 
The partial wave analysis has been performed 
for quasiparticle scattering off a vortex in a superconductor.\cite{cleary}
Though the phase shifts were obtained approximately, 
it is clear from the analysis that they are not all zero.
Using the formal relationship between the phase shift and the friction,
\cite{suhl,tt,chen,bonig}
the extended states have a contribution to the vortex friction,
in accordance with our results.

It should be emphasized that the logarithmic
divergence comes from the interplay between  the 
divergence in the density of states and the off-diagonal potential scattering.
We can consider a situation where we physically create a pinning center to
trap the  vortex and guide its motion. In such a case the vortex
has a diagonal potential. 
If the scattering is dominated by the diagonal potential, e.g., by the
trapping potential $U_0$, an additional factor coming from $|u_\alpha|^2 
- |v_\alpha|^2$ will remove this logarithmic divergence.
The friction on the physical trapping potential 
will be finite even above $T_c$ without the vortex,
as indicated in Refs. \onlinecite{suhl,tt,chen} and \onlinecite{bonig},
though the total transverse force disappears because $\rho_s =0$.
This again shows the sensitivity of the vortex friction to details.

The vortex friction from extended states exists 
for both clean and dirty superconductors at finite temperatures. 
Close to the transition temperature, it scales linearly with the superfluid
density,
and is exponentially small when $ T << \Delta_\infty$.
For intermediate temperatures $T \sim  \Delta_\infty$, 
using $\xi_0 \sim \hbar^2 k_F/m\Delta_\infty$ and
$N(0) = mk_F/\pi^2\hbar^2$,  $\eta 
\sim   L \hbar N(0) \Delta_\infty^2/k_BT $.  
When the impurity potential is nonzero, there is an additional
contribution to the friction, to be discussed in the next section.

We mention here that there is another type of low-lying excitation,
phonons, which may lead to an additional contribution to vortex friction.
This type of excitation can be described by the phase dynamics of 
the gap function $\Delta$, and has been ignored here by the assumption
of an adiabatic following up
of the gap function to the vortex coordinate.
Based on general considerations 
we expect that the phonon contribution is super-Ohmic.\cite{niu} 
Hence, it is asymptotically weaker 
than the (sub)Ohmic damping contribution from
quasiparticle excitations discussed above and the core state contribution
to be discussed below.

The nonzero friction contribution from extended states found above
is in accordance with the linear response theory in nonequilibrium statistical
mechanics, where transport coefficients are 
related to the fluctuations near the equilibrium by the 
fluctuation-dissipation theorem.
The fluctuations are completely determined by the underlying Hamiltonian.
Here they are quasiparticles determined by
the Bogoliubov-de Gennes equation.

It should be pointed that the expression leading to vortex friction, 
Eq. (47) or (49), are absent in Refs. \onlinecite{volovik,kopnin}
and \onlinecite{otterlo}.
To obtain a finite friction in those work, 
a finite relaxation time needs to be inserted
into the denominator of the force-force correlation 
function,\cite{otterlo} 
or into the denominator 
of the Nambu-Gor'kov Green's function at a convenient point,\cite{kopnin}
which at the same time leads to the reduction of the total transverse force.
As discussed in Sec. II.B, such a procedure should be avoided.

\subsection{ Core v.s. Extended States  Transitions for Total Transverse
              Force  }

There are various ways to express the transverse correlation function in 
Eq. (36) or (52),  with emphasis on different aspects 
of the transition elements.
In Sec. III.C we have shown how to obtain the total transverse force from the
consideration of extended states. 
In this subsection we show that it can also be obtained from the
consideration of core states, even some combination of both types of states.

Explicitly, we may evaluate the transverse response directly from 
Eq. (36) or (52).
We will show that the total transverse force can be expressed as 
contributions from only core to core transitions, 
or from  core to extended states transitions.
For a  clean superconductor we  replace 
$\nabla_{x_0}\rightarrow -\nabla $. 
We define following symbols for the transition elements:
\bea
    \alpha(l) \alpha'(l') & = & - i \hbar 
         (f_{\alpha}-f_{\alpha'}) \frac{1}{2} \hat{z} \cdot \int d^3x d^3x'
         \nonumber \\
   & &  \Psi_{\alpha}^{\dag}(x') \nabla\Psi_{\alpha'}(x') \times
      \nabla \Psi_{\alpha'}^{\dag}(x)  \Psi_{\alpha}(x)  \, ,
\eea
which groups the transition elements into core state
to core state, core state  to extended states,  
extended state to core states, and 
extended state  to extended state transitions.
Here $l=c,e$ represents the core $c$ or extended $e$ states,
and $\alpha$ represents other indices: $k_z$, $\mu$.
For example, $\alpha(c)\alpha'(c)$ represents the elements in Eq. (52)
 when $\Psi_{\alpha}$ and $\Psi_{\alpha'}$ are both core states. 
More explicit examples will be given below.
>From Eq. (52) and (65), 
the summing over these transition elements as well as over
$\alpha$ and $\alpha'$  gives $ B $,
\be
  B = \sum_{\alpha,\alpha'}
  \left[ \alpha(c)\alpha'(c) + \alpha(c)\alpha'(e) 
       + \alpha(e)\alpha'(c) + \alpha(e)\alpha'(e) \right] \; .
\ee
First, we note that for a core state, the sum
of its transition elements to all other states is zero: 
\be 
  \sum_{\alpha'} \left[ \alpha(c)\alpha'(c) 
                      + \alpha(c)\alpha'(e)\right] = 0 \, .
\ee
In fact, we have already obtained Eq. (67) and 
 used this identity earlier in Sec. III.C from
Eqs. (52)-(55) to exclude the core state contribution
to the circulating current far away from the vortex core in Eq. (55). 
In Eq. (52) both extended and core state contributions are there.
Then we summed
over all the $\Psi_{\alpha'}$ to reach Eq. (53) or (54),
because they form a complete set. 
Thus in the last expression of Eq. (52) inside the  sum over $\alpha$,
if  $\Psi_{\alpha}$ is an extended state, its contribution to the
transverse  response is actually 
$\sum_{\alpha'} \left[ \alpha(e)\alpha'(c) + \alpha(e)\alpha'(e) \right]$, 
because all the $\alpha'$s have been summed over. The same procedure
applies if $\Psi_{\alpha}$ is a core state. In Eq. (55), we have shown that
the area integral can be converted into a line integral and we can
choose the loop large enough compared to the core size. If 
$\Psi_{\alpha}$ is a core state, its contribution to Eq. (55) is zero.
Thus we conclude the validity of Eq. (67).

With the aid of Eq. (67), the transverse correlation can be expressed in the 
following two additional forms:
\bea
   B & = & \sum_{\alpha,\alpha'} 
                 \left[ \alpha(e)\alpha'(c) + \alpha(e)\alpha'(e) \right]  \\
      & = & \sum_{\alpha,\alpha'} \left[ \alpha(e)\alpha'(e) 
                                      -  \alpha(c)\alpha'(c)\right] \; .
\eea
In reaching Eq. (69), 
we have used the identity $ \alpha(c)\alpha'(e) = \alpha(e)\alpha'(c) $.
Here Eq. (68) can be reduced to Eq. (55) after using the completeness of the
eigenfunctions, as discussed above. 
Next, we present more detailed discussions on the 
core-core and extended-extended transition element contributions.

To be specific, we consider the zero-temperature case.
For the core states, because of the topology
the energy is uniquely determined by $\mu$.
The only transition elements contributed to the transverse correlation
function $F_{\perp}$ are between states $\mu = \pm \frac{1}{2} $:
\bea
 & &  \sum_{\alpha,\alpha'} \alpha(c)\alpha'(c) \nonumber \\
  & & = - i 2 \hbar 
    \left( f_{\mu=-\frac{1}{2} } - f_{\mu' = \frac{1}{2} } \right)  
      \frac{1}{2} \hat{z}\cdot
      \int d^3x \int d^3x' \nonumber \\
  & & {\ }      \left(  \Psi_{-\frac{1}{2}}^{\dag}(x') 
            \nabla_{x_0}\Psi_{ \frac{1}{2}}(x')\times
            \nabla_{x_0}\Psi_{\frac{1}{2}}^{\dag}(x)  
                        \Psi_{-\frac{1}{2}}(x) \right) \; . 
\eea
The additional factor 2 accounts for the transition from the 
$\mu=1/2$ state to the $\mu'= - 1/2$ state,
which gives the identical contribution.
The transition element in Eq. (70) may be expressed as
\bea
   & &  \int d^3x \int d^3x' \nonumber 
        \left(  \Psi_{-\frac{1}{2} }^{\dag}(x') 
               \nabla_{x_0}\Psi_{\frac{1}{2}}(x')\times
       \nabla_{x_0}\Psi_{\frac{1}{2}}^{\dag}(x) 
                  \Psi_{-\frac{1}{2}}(x)\right)\nonumber \\
   & & =  - i \hat{z}\delta_{k_z,k_z'}
       \frac{ \left(a_{-\frac{1}{2},\frac{1}{2}}    
                  + b_{-\frac{1}{2},\frac{1}{2}}       \right)
              \left(a_{-\frac{1}{2},\frac{1}{2}}^{\ast} 
                  + b_{-\frac{1}{2},\frac{1}{2}}^{\ast} \right) }
            { 2 \left(E_{-\frac{1}{2}} - E_{\frac{1}{2} }\right)^2 }  
           \; ,      \nonumber 
\eea
where
\bea
   a_{-\frac{1}{2}, \frac{1}{2}} & = & - \int_0^{\infty} r dr |\Delta|'_r 
        \left[ \hat{f}_{+,-\frac{1}{2}}^{\ast}(r) 
               \hat{f}_{-,\frac{1}{2}}(r)  \right. \nonumber \\
    & & \left. + \hat{f}_{-,-\frac{1}{2}}^{\ast}(r) 
               \hat{f}_{+,\frac{1}{2}}(r) \right] \; , \nonumber 
\eea
and 
\bea
   b_{-\frac{1}{2},\frac{1}{2}} & = & - \int_0^{\infty}  dr |\Delta|  
     \left[ \hat{f}_{+,-\frac{1}{2}}^{\ast}(r) \hat{f}_{-,\frac{1}{2}}(r)
           \right. \nonumber \\
  & & \left. - \hat{f}_{-,-\frac{1}{2}}^{\ast}(r) \hat{f}_{+,\frac{1}{2}}(r) 
               \right] \; ,  \nonumber 
\eea
which follow the definitions in Eqs. (A10)-(A15) in the Appendix.

Now we evaluate Eq. (70) explicitly. 
For the deep core states in clean superconductors 
$E=-\mu \epsilon_0$ with $\epsilon_0$ the core level spacing, 
$ f_{\mu=-\frac{1}{2} } = 0 $ and 
$ f_{\mu=\frac{1}{2} } = 1 $. 
The relation between the energy of core states and the
quantum number $\mu$ in our case is different from the
one in Ref.\onlinecite{bardeen} in sign because we are considering a vortex
with positive vorticity. 
The wave functions for deep core states take the form
\be
   \Psi_{\mu }^{\dag}(x ) \approx  \frac{1}{2}
        \sqrt{ \frac{ k_F }{ \xi_0} }
            \left( \begin{array}{c} 
        e^{ i ( \mu + \frac{1}{2} )  \theta} 
          J_{ \mu + \frac{1}{2} }(k_F \; r ) \\
           e^{ i (\mu - \frac{1}{2} ) \theta} 
          J_{  \mu - \frac{1}{2} }(k_F \; r )  
                    \end{array} \right) 
      e^{ - {r}/{\xi_0} }  \; .
\ee
This leads to $ a_{-1/2,1/2}(E,E') \approx  0  $ and 
$  b_{-1/2,1/2}(E,E') \approx \Delta_{\infty}/\xi_0 $.

Using Eq. (71), it is straightforward to show
\bea 
   & & \hat{z}\cdot
   \sum_{\mu,\mu'} ( f_{\mu} - f_{\mu'} ) \int d^3x \int d^3x'  
       \left(  \Psi_{\mu}^{\dag}(x') \nabla \Psi_{\mu'}(x')
                   \times \right.    \nonumber \\
   & & \left. \nabla\Psi_{\mu'}^{\dag}(x)  \Psi_{\mu}(x) \right)
    =  i k_F^2 \, .                  \nonumber 
\eea
Therefore,
\be
    \sum_{\alpha,\alpha'} \alpha(c)\alpha'(c) 
  =  \frac{\hbar k_F^2}{2}  
  =  B  \; .
\ee
The last equality is due to the fact that
in 2D, the electron density $n_e = k_F^2/2\pi$.
The additional factor $1/2$ accounts for the pairing.
The conclusion is that
at zero temperature the sum of the core-core state transitions alone 
gives rise to the total transverse force.
It corresponds to the fact that the core-core state transitions
are a local and differential form of the geometric phase, 
and the Berry phase is the  global and integral form.
In the next section we will show that 
Eq. (72) is unchanged in the presence of impurities.

Here we wish to point out an interesting feature explicitly manifested 
in Eq. (72):
The transverse response is insensitive to the size of the system, 
because the core states are exponentially localized.
This implies that 
the thermodynamic limit is not important for the total transverse force.
We attribute this feature 
to the topological constraint on the transverse response, corresponding to 
the well-known fact that 
the Berry phase exists for a discrete energy spectrum.

There are two more interesting results which follow Eq. (72).
At zero temperature, using Eqs. (69) and (72), we have
\be
   \sum_{\alpha,\alpha'} \alpha(e)\alpha'(e) = 2 B  \;  .
\ee
This implies that for a fermionic superfluid, the sum of
all extended state transitions lead to twice that of the total transverse
force. 
Combining Eq. (67) with Eq. (72), we have
\be
   B  = -  \sum_{\alpha,\alpha'} \alpha(e)\alpha'(c)  \; ,
\ee
which shows that the core-extended state transitions
can also be used to calculate the total transverse force.
We believe that this property has been explored before in the case 
considering the contributions from states whose energies 
are around $\Delta_{\infty}$, the interface between the
core and extended states.\cite{hk}

In the literature, after a transverse response
equivalent to Eq. (36) or (52) was reached, 
it had always been assumed that only one core-to-core state
transition contributes.\cite{kopnin,otterlo} 
However, as we have found out, core to extended state
transitions are of the same order.
The above discussions show that
there are many equivalent ways to compute the total transverse force.
Because of the topological constraint,
the total transverse force can even be evaluated by partial summations of 
the transition elements, expressed by Eqs. (68), (69), (72), (73), and (74).
This is completely different from the computation of 
the longitudinal force (the friction)
where a partial summation contributes only a part of the total friction.
The demonstration in this subsection suggests that 
the alleged cancellation between the core spectrum flow contribution and
the Berry phase counting is a consequence of the combination 
of double counting, treating core and extended contributions to
the transverse force as different quantities, 
and the misuse of the relaxation time approximation  
in the force-force correlation functions.

To briefly summarize this subsection,
we have formulated the transverse response in terms of transitions
between core state or between core and extended 
states. Equivalently, we have also formulated it in terms of 
a summation over the extended state contributions. 
In a clean and neutral superfluid at finite temperatures, 
the total transverse force is given by the product of 
the superfluid number density,
the Planck constant $h$, and the vortex velocity, 
though  the vortex friction exists.

\section{ Effects of Impurities}

\subsection{ No Effect on Total Transverse Force }

The presence of impurities is unavoidable in superconductors. 
In this subsection we consider this realistic situation of 
the influence of the impurity potential to the 
transverse force on the  moving vortex.   
In Sec. IV.B we have shown that
the transverse correlation function can be evaluated by either considering
the extended states or by considering only the core states
in a clean superconductor. 
The same also holds in the case with impurities, as we will demonstrate below. 
We first give a formal demonstration from the
counting of individual state contributions,
then explicitly consider the core state transitions, 
to pave the way for the core state contribution to vortex friction.
The robust conclusion is that random impurities do not affect 
the total transverse force.

It is more convenient 
to  change the gradient from $\nabla_{x_0}$ to $\nabla$ 
when there is an impurity potential $V(x)$ involved.
Applying  Eqs. (42) and (43) to Eq. (52), 
the transverse correlation becomes
\bea
  B 
  & = & - i \sum_{\alpha,\alpha'} \int d^3x \int d^3x'  
      ( f_{\alpha} - f_{\alpha'} )  \times  \nonumber \\
  & &      \frac{\hbar}{2} \hat{z}\cdot
       \left(  \Psi_{\alpha}^{\dag}(x') \nabla \Psi_{\alpha'}(x')
        \times  
       \nabla\Psi_{\alpha'}^{\dag}(x)  \Psi_{\alpha}(x) \right)  \nonumber \\
  & & + \sum_{\alpha,\alpha'} 
   ( f_{\alpha} - f_{\alpha'} )  
   (E_{\alpha} -E_{\alpha'})^{-2} \frac{ \hbar}{2} \hat{z}\cdot  \nonumber \\
  & &  \int d^3x \int d^3x'  
   \left(  \Psi_{\alpha}^{\dag}(x')\nabla{\cal H}'(x')\Psi_{\alpha'}(x')
        \times  \right.   \nonumber \\
  & & \left. \Psi_{\alpha'}^{\dag}(x) 
      \nabla{\cal H}'(x) \Psi_{\alpha}(x) \right) \; .
\eea
Other terms are identically equal to zero after summing over $\alpha, \alpha'$.

We first show that the second term in Eq. (75) is zero after the
impurity average.
To be concrete,
we expand the wave function $\Psi_{\alpha}$ 
in terms of  eigenfunctions of $\bar{\cal H}_0$, 
$\{ \phi_{\gamma} \}$, 
\be
  \Psi_{\alpha}= \sum_{\gamma} 
       \chi_{\alpha\gamma} \; \phi_{\gamma} \, .
\ee
Here $\chi_{\alpha\gamma} = a_{\alpha\gamma}e^{i\varphi_{\alpha\gamma}}$, and
$a_{\alpha\gamma}$ and $ \varphi_{\alpha\gamma}$ 
are the modulus and phase of the expansion coefficients. 
We remind the reader that
$\bar{\cal H}_0$ has dependence on
the impurity potential because
it  includes the smooth part of the  self-consistent
potential $\Delta$.
Because of the normalization requirement,
the coefficients $\{ a_{\alpha\gamma} e^{i\varphi_{\alpha\gamma} } \}$
form a unitary matrix,  
\[ 
   \sum_{\gamma} a_{\alpha\gamma}^2 = 1 \; , \; 
   \sum_{\alpha} a_{\alpha\gamma}^2 = 1  \; .
\]
The second term of Eq. (75) is now 
\bea
 & &\int d^3x  \int d^3x' 
   \Psi_{\alpha}^{\dag}(x') \nabla{\cal H}'(x')  \Psi_{\alpha'}(x') \times 
        \nonumber \\ 
 & & {\ } \Psi_{\alpha'}^{\dag}(x) \nabla{\cal H}'(x) \Psi_{\alpha}(x) 
        \nonumber \\
 & & = \sum_{\gamma, \gamma', \gamma_1, \gamma_1' } 
   a_{\alpha \gamma}     a_{\alpha' \gamma'}
   a_{\alpha' \gamma_1'} a_{\alpha \gamma_1}
   e^{- i\varphi_{\alpha \gamma}   + i\varphi_{\alpha' \gamma_1' } 
      - i\varphi_{\alpha' \gamma_1'} + i\varphi_{\alpha \gamma }} 
            \times \nonumber \\
 & & {\ }  \int d^3x \int d^3x' 
   \phi_{\gamma }^{\dag}(x') \nabla{\cal H}'(x')  \phi_{\gamma'}(x') \times  
         \nonumber \\
 & & {\ } \phi_{\gamma_1'}^{\dag}(x) \nabla{\cal H}'(x) \phi_{\gamma_1}(x) \, .
\eea
>From the random-matrix theory,\cite{random} 
the phase $\{ \varphi_{\alpha \gamma } \}$ are random numbers.
The impurity average makes $\gamma = \gamma_1$ and  
 $\gamma' = \gamma_1'$.
Under the assumption, i.e.,
the core size much larger than the average distance between impurities,
the averaging over impurities restores 
the homogeneity and isotropy of the spatial space.
This implies that all the odd power of  $\nabla{\cal H}'$
will be averaged to zero.
Since each term in Eq. (77),
\bea
 & & \int d^3x \int d^3x' 
    \phi_{\gamma}^{\dag}(x') \nabla{\cal H}'(x') \phi_{\gamma'}(x') \times  
    \phi_{\gamma'}^{\dag} (x) \nabla{\cal H}'(x) \phi_{\gamma}(x) \nonumber \\
 & & = \hat{z} \int d^3x \int d^3x' \left\{
       \left( \phi_{\gamma}^{\dag}(x') \nabla {\cal H}'(x')  
              \phi_{\gamma'}(x') \right)_x    \times  \right. \nonumber \\
 & & {\ }\left( \phi_{\gamma'}^{\dag}(x) \nabla {\cal H}' (x) 
                 \phi_{\gamma}(x)   \right)_y 
                      - \left( \phi_{\gamma}^{\dag}(x') 
   \nabla {\cal H}'(x') \phi_{\gamma'}(x')  \right)_y \times 
         \nonumber \\
 & & {\ } \left. \left( \phi_{\gamma'}^{\dag}(x) \nabla {\cal H}' (x) 
                  \phi_{\gamma}(x) \right)_x  \right\}   \; ,  \nonumber 
\eea
consists of the elements of odd power of $x$ and $y$ 
components of $\nabla{\cal H}'$, the second term of Eq. (75) is zero after
the impurity average.

The average transverse correlation is then, following the same procedure from
Eqs. (52)-(55),
\be
   B =  \oint_{|x - x_0| \rightarrow \infty }
            d{\bf l}\cdot{\bf j} \; .
\ee
Here the average current
\bea
    {\bf j} & = & - < \frac{i\hbar}{2} \sum_{\alpha} 
  \left[ f_{\alpha} u_{\alpha}^{\ast} \nabla u_{\alpha} 
   + (1-f_{\alpha}) v_{\alpha} \nabla v_{\alpha}^{\ast} \right] > + c.c. 
       \nonumber \\
    &  = & - \frac{i\hbar}{2} \sum_{\alpha\gamma } < a_{\alpha,\gamma}^2 > 
  \{  [ f_{\alpha} u_{\gamma}^{\ast} \nabla u_{\gamma}  \nonumber \\ 
    & &  + (1-f_{\alpha}) v_{\gamma} \nabla v_{\gamma}^{\ast}  ] 
      + c.c.  \}   \; , \nonumber  
\eea
and $< ... > $ stands for the impurity average 
over the expansion coefficients. 
In the limit $ r \rightarrow \infty$, we have
\bea
   {\bf j} & = & - \frac{i\hbar}{2} \sum_{\alpha,\gamma } 
     < a_{\alpha,\gamma}^2 > 
     \left\{ \left[ f_{\alpha} |u_{\gamma}|^2   \right. \right.   \nonumber \\
   & & \left. \left. 
     + (1-f_{\alpha}) |v_{\gamma}|^2 \right] + c.c. \right\} \nabla \theta
       \nonumber 
\eea       
At zero temperature, 
we have \bea
  & & \sum_{\alpha, E_{\alpha} > 0 } \sum_{\gamma } < a_{\alpha \gamma}^2 >
       | \bar{v}_\gamma (|x-x_0| \rightarrow \infty )|^2 \nonumber \\  
  & & =  \sum_{\alpha, E_\alpha > 0 } 
       < |v_\alpha(|x-x_0 | \rightarrow \infty )|^2 >  
         \nonumber  \\ 
  & & =  \rho_0  \; . \nonumber 
\eea 
The above second equality is the Anderson theorem, in that nonmagnetic
impurities do not affect the density of states near the Fermi surface, 
hence there is no effect on the superconducting transition temperature.
We will come back to this point after the discussion of the impurity effect
on vortex friction. 
This result may also be reached from the envelope 
wave function argumentation.\cite{degennes}  
Therefore  Eq. (55) also
holds in the presence of impurities.
Thus we have shown in general that 
the transverse correlation is not influenced by impurity potentials.
Physical understanding of this result is straightforward:
There is no average circulation current associated with impurity potentials.

Next, the insensitivity of the total transverse force
to random impurities will be illustrated by a different demonstration. 
We evaluate core state transitions with impurity potentials,
a part of the first term in the right-hand side of Eq. (75).
First, we will explicitly consider the total transverse force from core states 
contributions with a weak impurity potential. 
Because of the factor $(f_{\mu}-f_{\mu'})$ and the selection rule
in the transition elements, deep core states  are the ones 
making the main contributions to the total transverse force 
for temperatures well below $T_c$.   
When the impurities are weak enough, for deep core states $\Psi_{\mu} =
\sum_{\nu} a_{\mu\nu} e^{ i \varphi_{\mu\nu} } \, \phi_{\nu} $ 
the expansion coefficient  $a_{\mu\nu}$
is large only for the  neighboring states around $\mu$, which are
also deep core states. 
In a clean superconductor, 
core states are uniquely specified by the azimuthal number $\nu$.
Therefore, with  weak impurity potentials,
we may only consider deep core states close to the Fermi surface, 
and ignore the mixing of deep core-level states with the extended states
in Eq. (75).
 
Substituting Eq. (76) into Eq. (75) and using
Eq. (71) to calculate transition elements among
core states $\{ \phi_{\nu} \} $,  we have 
\bea
   & & \int  d^3x   \int d^3x'  
   \Psi_{\mu}^{\dag}(x') \nabla \Psi_{\mu'}(x') \times  
   \nabla\Psi_{\mu'}^{\dag}(x)  \Psi_{\mu}(x)  \nonumber \\
  & & = \sum_{\nu, \nu', \nu_1, \nu_1'} 
    a_{\mu \nu}a_{\mu' \nu'}a_{\mu' \nu_1'}a_{\mu \nu_1}
    e^{- i\varphi_{\mu \nu}   +  i\varphi_{\mu' \nu' } 
       -i \varphi_{\mu' \nu_1'} +  i\varphi_{\mu \nu_1 } } 
     \times \nonumber \\
 & & {\ } (\mp) i t_c^2 \delta_{\nu', \nu \pm 1 } 
                      \delta_{\nu_1', \nu_1 \pm 1 } \hat{z} \, .
\eea
Here we have used 
\[
  \int d^3 x \phi^{\dag}_{\nu}(x) \nabla \phi_{\nu'}(x) \approx
     i t_c \left( \pm \hat{x} + i \hat{y} \right) \delta_{\nu',\nu\pm 1} \; 
\]
with $t_c =  k_F/2$.
If $\mu$ and $\mu'$ are interchanged, there is a sign change in the 
right-hand side of Eq. (77).
Including the factor $ ( f_{\mu} - f_{\mu'} )$ and
summing over $\mu$ and $\mu'$, core-core state transitions become
\bea
   & & \sum_{\mu,\mu'} ( f_{\mu} - f_{\mu'} ) 
      \int d^3x \int d^3x'  \nonumber \\
  & & {\ } \hat{z}\cdot
       \left(  \Psi_{\mu}^{\dag}(x') \nabla \Psi_{\mu'}(x')
        \times  
       \nabla\Psi_{\mu'}^{\dag}(x)  \Psi_{\mu}(x) \right) \nonumber \\
  & & =   \sum_{\mu,\mu'} ( f_{\mu} - f_{\mu'} )
  \sum_{\nu, \nu', \nu_1, \nu_1'} 
   a_{\mu \nu}a_{\mu' \nu'}a_{\mu' \nu_1'}a_{\mu \nu_1} 
      \times \nonumber  \\
  & & {\ } e^{- i\varphi_{\mu \nu}   +  i\varphi_{\mu' \nu' } 
     - i\varphi_{\mu' \nu_1'}  + i\varphi_{\mu \nu_1 }} 
  \; 2(\mp) i t_c^2  \delta_{\nu',  \nu \pm 1 } 
                     \delta_{\nu_1', \nu_1 \pm 1 } \, .
\eea 
Because of the randomness in the phase factor of
$e^{- i\varphi_{\mu \nu}    + i\varphi_{\mu' \nu' } 
    - i \varphi_{\mu' \nu_1'} + i\varphi_{\mu \nu_1 } }$,
the dominant contribution comes from states 
$\nu=\nu_1, \nu'=\nu_1'$. Equation (80) becomes
\[
  \sum_{\mu,\mu'} ( f_{\mu} - f_{\mu'} )
  \sum_{\nu, \nu'} a_{\mu \nu}^2 a_{\mu' \nu'}^2
   2 (\mp) i t_c^2 \delta_{\nu', \nu \pm 1} \; .
\]

For a given distribution of $ a_{\alpha \mu}$ this summation can be
evaluated. For the purpose of demonstration, let us assume a 
simple distribution centered at $\alpha$:
$ a_{\mu \nu}^2 = 1/(2l+1)$ when $ | \nu - \mu |< l $,
 $ a_{\mu \nu}=0$ otherwise,  and consider only zero temperature.
Here $ 1 << l << N_c$, with $N_c$ being the total number of core states.
With this assumption,  each of the original
states is spread into a band of $2l+1$ states around it 
when impurities are present.
Equation (79) takes the value
\[
  \sum_{\mu,\mu'} ( f_{\mu} - f_{\mu'} )
  \sum_{|\nu - \mu| \leq l , |\nu' - \mu'|\leq l}  \frac{1}{(2l+1)^2}
  2 (\mp) i t_c^2  \delta_{\nu', \nu \pm 1 }  \; .
\]
We note that for the pair of  states $\mu=\mp 1/2, \mu'=\pm 1/2$ 
closest to the Fermi surface 
their contribution is reduced by a factor of
$ 1/ (2l+1)^2$. However, all the states within the energy shell
$ |E_{\mu}|  \leq l\epsilon_0$ near the Fermi surface
contribute now. In order to have a nonzero contribution at zero temperature,
we have $E_{\mu} E_{\mu'} < 0$, one above and one below the Fermi surface.
Because of the restriction of the band distribution and the selection for
$\nu'$ and $\nu$, we have an additional constraint 
on the $E_{\mu}$ and $E_{\mu'}$: $|E_{\mu}|, |E_{\mu'}| < l$.
The net contribution is  
\bea
 & & \sum_{\mu,\mu'} ( f_{\mu} - f_{\mu'} )
     \sum_{ |\nu - \mu| \leq l , |\nu' - \mu'|\leq l }  \frac{1}{(2l+1)^2}
      2 (\mp) i t_c^2 \delta_{\nu', \nu \pm 1 }   \nonumber \\
  & &  =  i k_F^2 / 2 \, , 
\eea
which is approximately the same value for the clean superconductor.
The factor of $1/2$ is due to the approximation for $t_c$
using same value for all transitions between $\mu$ and $\mu\pm 1$.
One can check that the above result also holds for 
Gaussian distribution of $a_{\alpha \mu}$. 
This completes the discussion of the first term in Eq. (75)
in the weak impurity potential limit.

Although at zero temperature in a clean superconductor
 only the core states closest to the Fermi
level contribute to Eq. (52) or (66), the transition elements of 
other states are not small. Their contributions
cancel each other completely. 
With impurities,  more states than those
closest to the Fermi level give contributions to the transverse response.
These contributions from other core states restore the
transverse response to its original value of a clean superconductor.
In the calculations with the relaxation time approximation,
the reduction of the contribution from the two states closest to the
Fermi level has been taken into account.\cite{kopnin,otterlo} 
The contribution due to
other core states, which arises  after introducing impurities, has
not been included in those calculations.

Next we consider the dirty limit, and we will again make use of arguments 
in the random matrix theory.\cite{random}  
We assume that there is no mixing between the 
core and extended states.
In the weak impurity potential limit, it is not difficult to justify this 
assumption:
The band width in Eq. (81) caused by impurities is much smaller than $\Delta$.
In the dirty limit, the number of core states remains the same, since 
the energy gap away from the vortex core remains the same.
Hence, there is a conservation of the number of core states,
because of the topological nature of the vortex.
We also note that there 
is no degeneracy for the core states, in contrast to the extended states.
In additional, impurities do not cause an additional violation of 
time-reversal symmetry. 
For those reasons we do not expect that they would mix
two topologically distinct types of solutions, the core and extended states,
of the Bogoliubov-de Gennes equation.

With increasing impurity potential strength, eventually any  
core states $\Psi_{ \mu }$ in Eq. (79) consist of all the core states 
of $\bar{\cal H}_0 $, 
$ \Psi_{\mu} = \sum_{\nu} \chi_{\mu\nu}\, \phi_{\nu} $.
Here the summation $\sum_{\nu}$ runs over core states only.
The total number of core states 
does not change after introducing impurity potentials, because
the core-level spacing for $\phi_{\nu}$ only depends on the values of
$E_F $ and $\Delta_{\infty}$, the value of  $\Delta$ far from the core.
A specific approximate 
realization may still be in the form of the band distribution as 
given in Eq. (81), with $l \sim N_c$, the total  number of core states:
\be
  |\chi_{\mu\nu}|^2 = \left\{
          \begin{array} {lcl} \frac{1}{N_c - 2|\mu| } & , 
       & |\mu - \nu| < N_c/2 - |\mu| \\
                         0 & , & |\mu - \nu| >  N_c/2 - |\mu| \; .
           \end{array} \right.
\ee             
It is widely spread for deep core states $|\mu| << N_c/2$.
One may check that $\sum_\nu |\chi_{\mu\nu}|^2 = 1 $ and
$ \mu = \sum_\nu |\chi_{\mu\nu}|^2 \nu $. The latter 
corresponds to the requirement for the
energy spectrum 
$<E_{\mu} > = < \Psi_{\mu} | {\cal H}_0 |\Psi_{\mu} >
 =  \sum_{\nu} |\chi_{\mu\nu}|^2 E_{\nu} $. 
The condition of 
$ \sum_\mu |\chi_{\mu\nu}|^2 = 1 $ is only approximately satisfied:
we have found that $ \sum_\mu |\chi_{\mu\nu}|^2 \approx \frac{1}{2}
    \ln ( N_c^2  )/( (N_c/2)^2 - \nu^2  ) $ which gives
$\ln 2$ when $\nu << N_c/2$ and $\ln N_c $ when $\nu \sim N_c/2$.

In this limit, Eq. (79) becomes
\bea
  & & \sum_{\mu,\mu'} ( f_{\mu} - f_{\mu'} ) \hat{z}\cdot
   \int d^3x \int d^3x'       
       \left(  \Psi_{\mu}^{\dag}(x') \nabla \Psi_{\mu'}(x')
        \times \right.  \nonumber \\
  & & {\ } \left. 
        \nabla\Psi_{\mu'}^{\dag}(x) \Psi_{\mu}(x) \right) \nonumber \\
  & & =  \sum_{\mu,\mu'} ( f_{\mu} - f_{\mu'} )
       \sum_{\nu, \nu', \nu_1, \nu_1'} 
      \chi_{\mu \nu}^{\ast}   \chi_{\mu' \nu'} 
      \chi_{\mu' \nu_1'}^{\ast} \chi_{\mu \nu_1} \times  \nonumber \\
  & & {\ } 2 (\mp) i t_c^2 \delta_{\nu',\nu \pm 1}
                           \delta_{\nu_1',\nu_1 \pm 1} \, .
\eea
Its average value is  $ ik_F^2/2 $ at zero temperature,
approximately 
the same as in clean superconductor, by using the distribution function
given in Eq. (82).

We have gone into great detail to calculate the total transverse force
from core state transitions in the presence of impurities.
Indeed, there is a reduction in transition amplitude 
between any pair of neighboring states.
Nevertheless, the summation over all possible core state transitions
restores the total transverse force to its value in the clean limit.
Hence, the impurities have a negligibly small effect on the total 
transverse force
from both the core state transitions consideration and 
the extended state counting, though
the friction contributions are strongly affected by impurities, 
to be discussed in the next subsection.

In addition, we check the self-consistent condition  
with respect to $\Delta$ here and show that they are satisfied
for our choice of $\chi_{\mu\nu}$. 
Because in $\bar{\cal H}_0$ we have already
assumed that the profile of $\bar{\Delta}$ is the smooth part
of the self-consistent potential with impurity potentials included,
we need to make sure that the decomposition of the eigenfunction
does not introduce an extra term to the self-consistent potential.
We have  
\bea
   \Delta &  = & -g\sum_{\alpha}
        u_{\alpha} v^{\ast}_{\alpha}(1-2f_{\alpha} ) \nonumber \\
  & = & - g \sum_{\mu,\nu,\nu'} \chi_{\mu\nu}
           \chi_{\mu\nu'}^{\ast} u_{\nu}^0 v^{0\ast}_{\nu'}
    (1-2f_{\alpha} ) \nonumber \\
  & & - g \sum_{\alpha(e)} u_{\alpha} v^{\ast}_{\alpha}(1-2f_{\alpha}) \, ,
\eea
where $u^0, v^0$ are the components of core 
eigenfunctions  $\{ \phi_{\nu} \} $, and 
$(e)$ denotes the extended states.
In the last equation, the summation is split into
those of core states and extended states.
Using a distribution function $p$ for $\chi $ such as defined by Eq. (82),
 the average core state contribution to $\Delta$ 
is the same as the one calculated by using  $\{ \phi_{\nu} \} $, 
\bea
 & & -g <\sum_{\mu,\nu,\nu'}\chi_{\mu\nu}   
    \chi_{\mu\nu'}^{\ast} u_{\nu}^0 v^{0\ast}_{\nu'} (1-2f_{\mu} )>  
          \nonumber \\
 & & =  - g \sum_{\mu} u_{\mu}^0 v^{0\ast}_{\mu} (1-2f_{\mu} )  \; .
    \nonumber 
\eea
The extended states also need to be self-consistent.
We assume the impurity strength is strong enough to mix the core state
on the scale of $1/\xi$ but
too weak  to cause extended states distortion on the scale of $1/k_F$.
Then for  extended states the distribution of 
$ \chi_{\alpha\gamma}$ is a function $p(E_{\alpha} -E_{\nu})$.
It is straightforward to check that the extended state contribution to
$\Delta$ is the same as that of those calculated by using $\{ \phi_{\nu} \}$.

\subsection{ Impurity Contribution to Vortex Friction }

In the present of impurity potentials, there are two kinds
of contributions to the friction. The extended state 
contribution remains basically the same as
what we have discussed before.
The main difference is that the coherence length in Eq. (61) will change 
 when impurities are present.
 We will give a brief discussion here. 
Since the density of states remains unchanged,
we only need to evaluate Eq. (61) again. 
The transition elements are given by
\bea
   & & |\langle\Psi_\alpha|\nabla_{x_0} {\cal H}_0 |\Psi_{\alpha' } \rangle|^2
    \nonumber \\
  & & = \sum_{\nu,\nu', \gamma, \gamma'} 
    |\chi_{\alpha  ; \nu \gamma} |^2 
    |\chi_{\alpha' ; \nu \gamma'}|^2         \times  \nonumber  \\
    & & {\ } \left\{  \begin{array}{lcl} 
    \frac{\Delta_\infty^2}{2 \pi^2 k_F^2 } \; 
    \delta_{k_{z},k_{z}'} \delta_{\nu ,\nu'\pm1}
     & , & |\nu| < \xi  k_{\rho} \\
   0 & , & |\nu| > \xi  k_{\rho}      \; .   \end{array} \right. \; .
\eea
Here $\chi$ is the expansion coefficient in Eq. (76), $\xi$ is the coherence
length in the presence of impurities,
$l$ is the angular index of the state, and $E$ is the energy of the state.  
With a given distribution of  $\chi$
such that the
expansion coefficient is confined to the neighborhood of its
original energy, it can be 
shown that the extended state contributions
to the friction remain unchanged, except the change of the coherence length
of $\xi_0 \rightarrow \xi$.

In the presence of impurities, the core state energy levels are no longer
monotonically arranged according to azimuthal number or the angular momentum. 
In addition, it may become quasicontinuous under the impurity average.
The mixing caused by impurity potentials makes it possible to have 
transitions into energetically nonadjacent core states, as discussed in 
the previous subsection, as well as into energetically nearly degenerate 
core states. 
Thus the  core states can give another contribution to the vortex friction, 
similar to the residual resistance in a metal.

First, we consider the  weak impurity potential limit.
We assume the
effect of impurities is not so strong such that we can treat their 
influence on core states perturbatively.
Using  Eqs. (42) and (43)  we obtain
\bea
    & & \left|\int  d^3x \Psi_{\mu}^{\dag}(x)
      \nabla_{x_0}{\cal H}_0 \Psi_{\mu'}(x) \right|^2      \nonumber \\
 & & = 
   \left| - (E_{\mu'}-E_{\mu}) \int d^3x 
       \Psi_{\mu}^{\dag}(x)\nabla \Psi_{\mu'}(x) \right. \nonumber \\
 & & {\ } \left.  - \int d^3x \Psi_{\mu}^{\dag}(x)\nabla{\cal H}' 
      \Psi_{\mu'}(x) \right|^2 \, .   
\eea
For the leading-order contribution, we only need to
use the unperturbed $\Psi_{\mu}$ and $\Delta$. 
The first term in Eq. (86) will not give any contribution to the
dissipation because of the discreteness of $(E_{\mu'}-E_{\mu})$ and
the  factor
$ \delta(\hbar\omega - | E_{\mu} - E_{\mu'} |)$ in $J(\omega)$. 
After summing over $\mu$ and $\mu'$, this term  will
become terms of $\delta(\hbar\omega\pm\epsilon_0)$, which will
not give any dissipation. 
The contribution to the dissipation comes from the second term.

We may assume the impurity 
potential has a length scale small compared with the coherence length
so that we can describe it by a delta 
potential $V(x)=\sum_i V_0 \delta(x-x_i)$. We have
\[
 < \left| \int d^3x \Psi_{\mu}^{\dag}(x)
  \nabla {\cal H}' \Psi_{\mu'}(x) \right|^2 >
  = n_i (\pi\xi^2 L) V_0^2 \left(\frac{k_F}{\pi\xi^2 L}\right)^2  \; ,
\]
with the impurity concentration $n_i$.
Under this assumption we
will make connections to the normal-state
transport parameters.  Note that  for normal states the electronic 
transport relaxation time 
and the electron scattering cross section have the
following relations:\cite{mahan} 
\[
 \tau^{-1}_{tr}  = n_i  v_F \sigma_{tr}  \; 
\]
with
\[
  \sigma_{tr}  =  \int d\Omega \; ( 1 - \cos\theta)|V(\theta)|^2  \; .  
\]
Here
\[
  V(\theta) 
    = -\frac{m}{2\pi\hbar^2}\int d^3x V(x)e^{-i{\bf q}\cdot{\bf r}} \, 
\]
with ${\bf q}= {\bf k} - {\bf k'}$, $\theta$ is the angle
between ${\bf k}$ and ${\bf k'}$, and $v_F = \hbar k_F/m$.
For our choice of impurity potential, $\tau^{-1}_{tr}$ can be calculated, 
\[
   \tau^{-1}_{tr} =  n_i v_F
   \left(\frac{m}{2\pi\hbar^2}\right)^2  V_0^2 
   \int d\Omega \; ( 1 - \cos\theta) \, .
\]
We emphasize here that the electronic transport relaxation time $\tau_{tr}$ 
is directly determined by the Hamiltonian of Eq. (22). 

Expressed in $\tau_{tr}$, the spectral function now becomes
\bea
   J(\omega) & = & 2\hbar\omega \left(\frac{k_F L}{2\pi\epsilon_0 }\right)^2
 n_i (\pi\xi^2 L) V_0^2 \left(\frac{k_F}{\pi\xi^2 L}\right)^2
 \nonumber \\
 & =  & \omega \frac{3}{2} \frac{m n_e (\pi\xi^2 L)}{\tau_{tr}} \; .
 \eea
Here $\epsilon_0=\Delta^2_\infty /E_F$ is the core-level spacing,
and $E_F = m v_F^2/2$. 
In Eq. (87)
 $ k_F L /{ 2\pi \epsilon_0}$ is the approximate density of core states
near the Fermi surface. 
It appears with the factor $2\hbar \omega$ because 
\bea
 & & \sum_{\mu,\mu'}
  \delta(\hbar\omega -|E_{\mu} - E_{\mu'}|)|f_{\mu} - f_{\mu'}|
      \nonumber \\
 & & = \int d E_{\mu} d E_{\mu'}  
  \delta(\hbar\omega -|E_{\mu} - E_{\mu'}|)
   |f_{\mu} - f_{\mu'}|\times  n_c^2(E) 
             \nonumber \\
  & & = 2\hbar\omega n_c^{2}(E) \;   \nonumber
\eea
with $n_c(E) \approx k_F L/(2\pi\epsilon_0) $.
The scattering time $\tau_{tr}$ is linked to the
residual resistivity by
\[
  \rho = \frac{m}{n_e e^2 \tau_{tr}}
\]
and can be measured independently.

The above spectral function $J(\omega)$ 
gives the vortex friction in the weak impurity limit
\be
   \eta = \frac{3}{2} \frac{m n_e (\pi\xi^2 L)}{\tau_{tr}} \, .
\ee
It  has a simple interpretation. 
For a normal electron moving in the metal, the friction is simply 
$m/\tau_{tr}$. Equation (88) can be interpreted as that in the weak impurity
limit, the friction for a vortex is  the friction for 
each electron times the total number of electrons
inside the  core,  $n_e (\pi\xi^2 L)$.  

This vortex friction  increases with impurity concentration and strength.
We will show  that this increase eventually saturates in the dirty limit.
Using  Eq. (76),
we expand localized states $  \Psi$ in terms of $\{ \phi_{\nu} \}$, the 
of  eigenfunctions of ${\cal H}_0$.
\bea
   & & < \left|\int  d^3x \Psi_{\mu}^{\dag}(x)\nabla_{x_0}{\cal H}_0 
            \Psi_{\mu'}(x) \right|^2   >       \nonumber \\
   & & = \sum_{\nu, \nu', \nu_1, \nu_1'} 
    < \chi_{\mu \nu}\chi_{\mu' \nu'}\chi_{\mu \nu_1}
    \chi_{\mu'\nu_1'}  >  \times    \nonumber \\
   & & (E_{\nu} -E_{\nu'})(E_{\nu_1} -E_{\nu_1'})  \times \nonumber \\
   & & {\ }  \int d^3x \phi_{\nu}^{\dag}(x)\nabla \phi_{\nu'}(x)\cdot
    \int d^3x \phi_{\nu_1}^{\dag}(x)\nabla \phi_{\nu_1'}(x)  \nonumber \\
   & & = \sum_{\nu, \nu'}  |\chi_{k \nu}|^2 |\chi_{k' \nu'}|^2
    2 \epsilon_0^2 |t_c|^2 \delta_{\nu',\nu \pm 1 }  \; .
\eea
With the distribution function given in Eq. (82), the average value
\[
  <\left| \int d^3x \Psi_{\mu}^{\dag}(x)\nabla_{x_0} {\cal H}_0 
       \Psi_{\mu'}(x) \right|^2 >
        = \frac{4 \epsilon_0^2 |t_c|^2}{N_c} \, .
\]
Here the total number of core levels  is
\[
  N_c = 2 \Delta_{\infty}  \frac{k_F L }{ 2\pi\epsilon_0 } 
  = \frac{E_F}{\Delta_{\infty}}\frac{k_F L}{\pi} \, .
\]
Finally, the spectral function is  
\be
   J(\omega) =   {\hbar\omega}\left(\frac{k_F L}{2\pi \epsilon_0 }\right)^2
 \frac{4 \epsilon_0^2 |t_c|^2}{N_c}
 = \omega \frac{3\pi^2 }{8} \hbar n_e \frac{\Delta_{\infty}}{E_F}  L \; ,
\ee
which gives the friction per unit length
\[
  \eta =  \frac{3\pi^2 }{8} \hbar n_e \frac{\Delta_{\infty}}{E_F} \, .
\]
This result is similar to what is obtained in Ref.\onlinecite{bs}. Hence 
its microscopic base has been provided.
In the low-temperature limit, the magnitude of the 
vortex friction is smaller than the total transverse force by a factor of 
$\Delta_{\infty}/E_F$. 
 
In the above derivation, 
we have ignored the localization effect which suppresses the density of 
state, or the superfluid number density. 
We justify our assumption here.
There are three energy scales involved in the derivation of vortex dynamics, 
the Fermi energy $E_F$, the energy gap $\Delta_\infty$, and
the core level spacing $\Delta_\infty^2/ E_F$.
The effect of impurities on vortex dynamics is believed appreciable 
 at  $\tau_{tr} \Delta_\infty^2/\hbar E_F \leq 1 $,\cite{kopnin} 
and the equality of Eqs. (90) and (87)  suggests 
that the impurity starts to be effective at
$\tau_{tr} \Delta_{\infty}/\hbar \, (\Delta_{\infty}/E_F )^2 \sim 1 $.
They indicate that the impurity effect on vortex friction occurs at a rather
weak level, determined by the smallest energy scale in the problem.  
The dirty limit is given by 
 $ \Delta_\infty/E_F < \tau_{tr}  \Delta_{\infty}/\hbar < 1$.
The localization effect is only pronounced in the extremely dirty limit,
the localization regime,
when $\tau_{tr} E_F/\hbar \leq 1$.\cite{ma}
Because $\Delta_\infty/E_F << 1$, away from the localization regime 
the suppression of density is indeed negligible.
The unsuppressed electronic density applies,
 and the present results are valid well
into the dirty limit of the superconductors.

To summarize this section, 
we have shown that the total transverse force is insensitive to impurities
by two different methods, 
but the additional core contribution to the vortex friction arises.
For a weak enough impurity potential, a perturbative calculation
leads to the core friction proportional to the normal-state
resistivity. In the dirty limit the core friction contribution
saturates to a value determined only by the energy gap and the Fermi energy.

\section { Coupling to  Electromagnetic Field }

Now let us discuss a superconductor when the penetration depth
is finite but still much larger than the coherence length. 
The Lagrangian is  given by
\bea
   L_{BCS} &  = &   \sum_{\sigma}\psi^{\dag}_{\sigma}( x,\tau) 
      \left(  \hbar\partial_{\tau}- \mu_F - e A_0   + 
      \frac{1}{2m}\left(\frac{\hbar}{i} \nabla  \right. \right. 
               \nonumber \\
    & & {\ } \left. \left.  - \frac{e}{c}{\bf A}\right)^2
         + V(x)  \right) \psi_{\sigma}(x,\tau) \nonumber \\
    & & - g\psi^{\dag}_{\uparrow}(x,\tau) \psi^{\dag}_{\downarrow}(x,\tau)
          \psi_{\downarrow}(x,\tau) \psi_{\uparrow}(x,\tau ) \nonumber \\
    & & + \frac{1}{8\pi}(E^2 + B^2 ) + e A_0 n_0  \; ,
\eea
here $- e n_0$ is the charge density of the ionic background. 
The coupling to the electromagnetic field is in the usual minimum 
coupling form.
The fermionic degrees of freedom can be integrated out to give
\bea
   \frac{S_{eff} }{\hbar}& = &- Tr \ln G^{-1} +
     \frac{1}{\hbar g}\int_0^{\hbar\beta} d\tau \int d^{3}x|\Delta|^{2}
      \nonumber \\
& &     + \int_0^{\hbar\beta} d\tau \int d^{3}x
    \frac{1}{8\pi}(E^2 + B^2 ) +  e A_0 n_0 \; ,
   \eea 
with 
\be
    \ba{l}  ( \hbar \partial_\tau + {\cal H})
      G(x,\tau; x',\tau') = \delta(\tau-\tau') \delta^3(x-x') , 
    \ea
\ee
and
\be
   \ba{l}
   {\cal H}= \left( \begin{array}{cc} 
                     H & \Delta  \\
                    \Delta^{\ast} & - H^{\ast} 
                   \end{array} \right) \; .
   \ea
\ee
Here
\[ H =  - e A_0 +
 \frac{1}{2m}\left(\frac{\hbar}{i} \nabla - \frac{e}{c}{\bf A}\right)^2 
 -  \mu_F + V(x)\, , 
\]
\[
  H^{\ast}  = - e A_0 + 
  \frac{1}{2m}\left(\frac{\hbar}{i} \nabla + \frac{e}{c} {\bf A}\right)^2 
  - \mu_F + V(x) \, ,
\] 
and $c$ is  the speed of light in a vacuum.
Variation with respect to   $A_0$ and ${\bf A}$
gives 
\[
  \nabla\cdot{\bf E} = 4\pi e (n - n_0)  \; 
\]
and
\[
 \nabla\times {\bf B} - \frac{1}{c}\partial_\tau {\bf E}= 
 \frac{4\pi e}{c}{\bf j} \, 
\]
with $ {\bf E} = - \nabla A_0 + \frac{1}{c} \partial {\bf A}/\partial \tau$ and
     $ {\bf B} = \nabla \times {\bf A}$.
Here $e n$ and $e {\bf j}$ are the electric charge and current densities.
They should be obtained through the electronic Green's function.
In  the Lorentz gauge, the equations for 
$A_0$ and ${\bf A}$ from the above equations are,\cite{jackson}
adapted to the imaginary time here, 
\be
   \left[ \nabla^2 + \frac{1}{c^2 } \partial_{\tau}^2 \right] A_0 
    =  - 4\pi \; e ( n - n_0 ) \;  
\ee
and
\be
  \left[ \nabla^2 + \frac{1}{c^2 } \partial_{\tau}^2 \right] {\bf A} 
    =  - \frac{ 4\pi }{c} \;  e {\bf j}  \; . 
\ee
Assuming that for a static vortex at $x_v$ the vector potential is 
$\overline{\bf A}(x - x_v)$, then
for a slow moving vortex, the correction $\delta {\bf A} $ 
to the vector potential 
${\bf A}$ from $\overline{\bf A}= \overline{\bf A}(x - x_v(\tau))$
starts from second order in $\dot{x}_v(\tau)$
and can be ignored.
This can be directly demonstrated from Eq. (96) to the leading order
in $\delta {\bf A} $:
\[ 
   \nabla^2 \delta {\bf A} + \frac{1}{c^2} \partial_{\tau}^2  \overline{\bf A}
     = 0 \; . 
\]
The same is true for the scalar potential $A_0$.
For our purpose of keeping to linear order in $\delta \dot{x}_v$ 
we may use ${\bf A} = \overline{\bf A}(x - x_v(\tau))$.

Now we expand 
\[
  A_l (x - x_v(\tau)) = 
         \left( 1 
         + \delta x_v(\tau) \cdot \nabla_{x_0} \right) 
        \overline{A}_l (x - x_v(\tau)) \; ,
\]
with $l=(0,x,y,z)$.
The effective action for the vortex is
\bea
     \frac{S_{ eff } }{\hbar } & = &   \frac{1}{2} Tr (G_0 \Sigma' )^2 
        \nonumber \\
     & & + \frac{1}{\hbar g} \int_0^{\hbar\beta}\int d^3x 
          \delta x_v\cdot \nabla_{x_0} \Delta^{\ast}_0  \; 
                 \delta x_v\cdot \nabla_{x_0} \Delta_0 \; \nonumber \\ 
 & & + \int_0^{\hbar\beta} d\tau \int d^{3}x \; 
    \frac{1}{8\pi}(E^2 + B^2 ) 
 +  e  A_0  n_0  \; ,  
\eea
with
\bea
    \Sigma' & = & \delta x_v\cdot \nabla_{x_0} \left( \begin{array}{cc}
    - \frac{e}{2mc}\frac{\hbar}{i} (\overline{\bf A} \cdot \nabla)
         &  \Delta_0 \\
    \Delta_0 & - \frac{e}{2mc}\frac{\hbar}{i} 
      (\overline{\bf A} \cdot \nabla)
        \end{array} \right)  \nonumber \\
   & = & \delta x_v \cdot \nabla_{x_0} 
    {\cal H}  \; ,
\eea
where both ${\bf E}$ and ${\bf B}$ are substituted by the
stationary values calculated from
$A_0 = 0$ and ${\bf A}= \overline{\bf A}(x - x_v(\tau))$. 
Equation (97) is in the same form as Eq. (36).
There is then no change of the transverse response from the superfluid, 
reflecting the fact that the canonical momentum of the superfluid
is not changed by the coupling to the electromagnetic field.
A similar conclusion has also been reached by others 
phenomenologically.\cite{em-effect}

However, there are relativistic  corrections to the solutions 
(Eqs. (95) and (96) ) 
of the Maxwell equations due to the motion of the current and charge sources
associated with the vortex.
They are determined according to the Lorentz transformation of the
four vector formed by the scalar and vector potentials, 
or equivalently, the four vector by the charge density and current.
Those relativistic corrections give rise to additional terms  
in the action and can, in principle, contribute to the total 
transverse force
in vortex dynamics.
The relevant term in the effective action takes the form
\[
   \int_0^{\hbar\beta} d\tau \int d^{3}x  \; 
  \frac{e}{c} \delta \dot{x}_v \cdot \overline{\bf A}( x-x_v ) \; 
  [ n(x - x_v )  - n_0 ]  \; , 
\]
arising from the relativistic correction to the scalar potential,
the Aharonov-Casher phase.\cite{ac}
This contribution is due to the interaction between the moving magnetic
flux carried by a vortex to the electric charges of both 
conducting electrons and the background charges.
Since the charge neutrality condition is maintained 
in a superconductor, this contribution is zero, as also been
noticed in Ref.\onlinecite{stone}.
Hence, we do not need to consider it here.
Other relativistic corrections do not affect the total transverse force, and 
the rest of the terms have the same structure as the uncharged superconductor
with the replacement of Eq. (98) to Eq. (29). Therefore, all the
steps from Eqs (24)-(36) remain unchanged. 
We arrive at the same expressions (Eqs. (35) and (36)).
There are two differences, however.
First, the Bogoliubov-de Gennes equation
includes a vector potential, which can be served to generate a vortex, not 
by a rotation of superfluid. 
Second, we have now a compelling physical reason to
neglect the phonon (density fluctuation) mode compared to the
neutral case, because it is the plasma mode with a big energy gap. 

Let us discuss the effect of including the vector potential in 
the Bogoliubov-de Gennes equation to the final results. 
For the extreme type-II superconductor the vector
potential near the core is $A_{\theta} = \frac{1}{2} r h_0$, 
here $h_0$ is the magnetic field along the vortex line.
When the penetration depth is large, $h_0$ is small. For 
small $r$, when solving the vortex core structure, 
$\hbar c/2er > A_{\theta}$ and
we can safely ignore  $A_{\theta} = \frac{1}{2} r h$.
The core structure
is insensitive to the vector potential in the extreme type-II case.
Because the total transverse force can be expressed in core state
transitions, it is insensitive to coupling to electromagnetic
field. 
Equivalently, when expressing the transverse response in terms
of the summation over extended states, Eq. (55), we need the
large-$r$ behavior of the Bogoliubov-de Gennes equation.
When $r>>\lambda$, $A_{\theta}\rightarrow 0$, so that the
the coupling to the  electromagnetic
field will not influence the results in Eq. (55).
In the presence of impurities, the vortex friction is also
insensitive to the coupling to the electromagnetic field,
as noticed long ago.\cite{bs}

To summarize this section, 
in a charge neutral extreme type-II superconductor, the
vortex dynamics is the same as that in an uncharged BCS superfluid.
This is a known result, but we have sketched how to obtain it within
the present formulation.

\section{ Experiments }

\subsection{ Transport Measurement}

It had been assumed that the forces on a vortex could be 
extracted from  
transport measurements. Let us first review this apparently
plausible proposal 
and discuss  ideas which are crucial to the understanding 
of transport  measurements.
Considering our previous derivations, 
the  equation of motion of the ith  vortex  takes the
 form of the
Langevin equation similar to that of a
charged particle in the presence of a magnetic field:
\be
  0 = q_v  \rho_s h \; ( {\bf v}_s - \dot{\bf r}_i )
    \times \hat{z} - \eta  \dot{\bf r}_i + {\bf F}_{pin} + {\bf f}
  + \sum_j {\bf F}_{ij} \; .
\ee
Here $q_v = \pm 1$ represent different 
vorticity.
The total transverse force $ - q_v\rho_s h    \dot{\bf r} $ and
viscosity   $\eta$ are the ones we have calculated in the previous
sections. In addition, there are a
fluctuating force ${\bf f}$  related to the viscosity by the 
 fluctuation-dissipation theorem,  a pinning force ${\bf F}_{pin}$, 
and we should also include the
forces due to other vortices $ {\bf F}_{ij}$ because of vortex interaction.
Here we have explicitly written out the external current term in Eq. (99),
though in a real situation its effect is always through
the rearrangement of vortices in the superconductor.
The motion of vortices is a genuine many-body problem. 
A general exact solution does not exist.

Equation (99) may be solved after a drastic simplification by
 ignoring the pinnings.
This is equivalent to the situation that a perfect 
vortex lattice is sliding through the sample. Together with 
with the Josephson relation, we can determine longitudinal and
transverse resistivity for superconductors.
The Hall angle, defined as $ \theta_{Hall} = \tan^{-1}(\rho_{xy}/ \rho_{xx})$,
is nearly 90 degree for almost all situations. 
However, in transport measurement, most of the samples
show a small Hall angle  and some show a sign
change in the Hall angle upon entering the superconducting state.
This simplified model certainly disagrees with experiments. 

Now let us consider whether or not
this simplification can be made by 
considering the magnitude of pinning.
The equation of motion of a vortex in a superconductor, Eq. (99),
has the form of a particle with zero mass in a strong magnetic field.
Because the kinetic energy is zero, it is always confined to a
local energy minimum in space,
formed by vortex interactions and pinning potentials. 
An applied current tilts the potential and the vortex moves by thermal 
activation.
If the applied current is so large that no minima due to pinning and
vortex interaction exist, then indeed we expect a large Hall angle. 
However, in order to have a truly free vortex flow, the current
should be large enough to overcome the largest pinning potential, the
edge pinning. The current needed is on the order of 
$10^{8}$ A/cm$^2$, which  is too large to be relevant
to experiments. Therefore, in the real experiments, the
vortices must be helped by their many-body interactions to
overcome this energy barrier. We need to consider transport measurement by 
solving the lattice structure formed by vortices and by what mechanism
their transport is made possible. 
It has been quantitatively suggested that vortex many-body effects can be 
responsible for the Hall effect,
in agreement with recent experimental indications.\cite{hall}

\subsection{ Direct Measurement of Total Transverse Force }

The total transverse force on a moving vortex in thin YBCO films has been
directly measured via a mechanical device.\cite{zhu} This experiment used a 
small vibrating magnet mounted above the center of a 
superconducting film to generate moving vortices in the film. The 
vortices follow the motion of the magnet for samples with less twin boundaries.
The experiments were performed  on those samples.
The force was measured by measuring the motion of the superconducting
film in response to the vibration of the magnet. 
The experimental results have provided a qualitative confirmation 
of the insensitivity of the total transverse force to impurities.

\subsection{Measurement of Friction}

In a rf resistance measurement, 
the vortices are moving around their local minima, rather than
over potential barriers in a DC measurement. In such a case, it is possible
to observe intrinsic friction after using a potential to
describe the periodic vortex interaction and making some assumptions
about the pinning. The rf resistance was analyzed early with a
vortex dynamics model without transverse force.\cite{rf}
In order to compare with our theory, the total transverse force needs 
to be included. 
We will not go into any detail other than to suggest this possibility.

\section{ Conclusions  }

We summarize here what we have achieved in the present paper. 
With respect to the microscopic derivation,
we have developed an influence functional formulation started from the
BCS theory.
This formulation has allowed us to discuss several difficult questions
regarding vortex dynamics.
One question has been whether the total transverse force originates from
core states, extended states, or from both. 
This question is unique to a
fermionic superfluid because of the vortex  core structure.
We have shown that the total transverse force 
can be calculated equivalently by considering exclusively 
transitions between  core states,
by transitions between core and extended states, or by counting
contributions from extended states.
The total core-state transition contribution to 
the total transverse force is shown not to be affected 
by impurities when calculated 
by using random matrices instead of the relaxation time approximation.

On the thermodynamics and statistical mechanics level, 
we need to consider the increase
of the superfluid kinetic energy associated with
the increase of superfluid momentum due to the vortex motion.
This kinetic energy needs to be provided from somewhere. 
If there are no normal fluid and no impurities, 
this kinetic energy is provided  from the  work
done by the external trapping potential on the vortex. 
When either the normal fluid or impurities, or both, are present, 
there is a question whether or not a vortex can extract the internal energy 
from the normal fluid or substrate, which carries entropy,
and can transfer it into  kinetic energy of the superfluid, 
which carries no entropy. 
If not allowed, the increase of superfluid kinetic energy
due to vortex motion can only be provided by an external force
and the transverse force on a vortex cannot be reduced by the normal fluid
or random impurities. 
We have discussed this question 
and demonstrated that thermodynamics gives a powerful
constraint on phenomenological models of vortex dynamics: 
The total transverse force cannot be reduced.

We have located the source for contradicting
theoretical results in the two pictures:
the use of the relaxation time approximation
in the force calculation. This problem is rather subtle.
It is well known that the relaxation time approximation
has been used successfully in some applications, 
particularly in calculations of conductivity or mobility, where the average 
velocity is computed under a given driving force, 
through velocity-velocity correlations.
However, the special feature of 
vortex dynamics is that it belongs to the same category as 
resistivity or friction formulas in transport theory, 
where the average force is computed  with a given velocity.
The direct calculation of resistivity is known to be difficult.
To derive vortex dynamics microscopically, 
the vortex velocity-velocity correlation is not calculable directly, 
because the effective vortex Hamiltonian is unknown
and is precisely what we are looking for.
We are forced to abandon the usual approaches of the Nakano-Kubo type, 
and to tackle the problem from the difficult side.
Nevertheless, in the DC limit, the transverse force on a moving
vortex can be calculated from the force-force correlation function, 
in analogy to a DC resistivity formula.
This limit makes the relaxation time approximation invalid, 
because a significant part of the frequency dependence is lost, and
the common way of introducing the relaxation time approximation 
by substituting
 $i\omega\rightarrow i\omega + 1/\tau$ 
requires the correct frequency dependence. 
For example, in transport theory, the relaxation 
time approximation is used in an AC
conductivity formula, then taking the DC limit subsequently.
In addition, the relaxation time approximation 
in a force-force correlation function
is always erroneous. 
With an exactly solvable model, we have shown that when
the relaxation time approximation is used  in a DC resistivity formula, 
it leads to results violating fluctuation-dissipation theorems.

Introducing the relaxation time approximation, even done correctly, 
is not a necessary step in obtaining dissipation. 
One of the goals of nonequilibrium statistical mechanics
is to compute various transport coefficients, including the 
relaxation time, for a given Hamiltonian system.
In a Hamiltonian system, dissipation appears after 
irrelevant degrees of freedom are integrated out.
What determine dissipation are quantities like temperature
and strength of impurity potentials, as well as 
the density of state of low-frequency modes of irrelevant degrees of freedom.
In the present paper, irrelevant degrees of freedom are the fermionic
quasiparticles.
When those quasiparticle degrees of freedom are eliminated,
one obtains the vortex friction.
The friction formula obtained here 
follows the one used in  dissipative quantum dynamics,\cite{leggett} 
where it has been explicitly shown that the friction obtained
by eliminating  irrelevant degrees of
freedom is equivalent to an evaluation of the random  
force-force correlation function.
It also corresponds to the familiar Fermi Golden rule 
for dissipation.
We believe that a rather detailed study of vortex dynamics 
based on the BCS theory have been presented here, with 
the key issue of the sources for the vortex friction.
We have shown that the vortex friction can come from two contributions:
At finite temperatures, the finite population of quasiparticles
above and quasiholes below the energy gap give rise to a friction  
which diverges logarithmically at low frequency;
The nonmagnetic impurities give rise to an extra friction
which saturates to a value independent of
the normal-state resistivity in the dirty limit.
This core state contribution corresponds to the phenomenological value
obtained in Ref.\onlinecite{bs}.
We have also considered the effect of  
coupling to the electromagnetic field and have found that it does 
not change the neutral superfluid conclusions 
when the superconductor is  charge neutral, consistent with
earlier phenomenological treatments.

Finally, we expect that the method developed here to 
formulate the vortex dynamics in an s-wave superconductor
will find applications in other systems represented by
dynamics of collective variables, such
as vortex dynamics in d-wave superconductors, 
fission and fusion in atomic nuclei, and even the quasiparticle 
dynamics in quantum Hall systems.

{\ }

\noindent Acknowledgments

We are very grateful to David Thouless for many valuable discussions 
at various stages of the work, 
which have deepened our understanding of physics, 
sharpened our arguments, 
and lead to a better presentation of our results.
Discussions with Tony Leggett on several issues of dissipation
greatly clarify our thoughts.
We appreciate the communications from Mike Stone
informing us of his work and helping us to understand
the relevant work better. 
We thank R. K\"ummel for the detailed discussion
of Eqs. (38) and (49).
Discussions with Jung Hoon Han and Qian Niu are also acknowledged. 
We also thank the hospitality of the Department of Physics and 
the Institute for Nuclear Theory (P.A.) at the University of Washington, where
a part of the work was done. 
This work was financially supported by the Swedish Natural Science Research
Council (NFR).

{\ }

\appendix 

\section{ Divergent Overlap Integrals and Vortex Friction }
 
There have been some questions above the implications of Eqs. (38) and (39),
repeatedly raised by referees as well as by others during private discussions,
in particular on the diverging nature of the overlap integrals
on the right-hand side of Eqs. (38) and (39), when the energy difference 
between two eigenfunctions vanishes.
This question may have already been addressed in the literature.
Nevertheless, we believe it is helpful
to give it an explicit discussion in the present context.

We note that Eqs. (38) and (39) are exact consequences
of the fact that the Hamiltonian ${\cal H}_0$ is the function of the 
parameter $x_0$.
To make the connection to the scattering problem of quasiparticles scattered
off a vortex, the thermodynamic limit must be taken first to allow
the existence of the continuous spectrum.
This implies that it is appropriate to use the Dirac delta function
normalization for extended states,
\setcounter{equation}{0}
\be
   \Psi_{\alpha}  =
   \left( \begin{array}{c} u_\alpha (x) \\ v_\alpha (x) \end{array} 
      \right) =  \frac{e^{ik_z z} }{\sqrt{L}} 
             \frac{ e^{i\mu\theta } }{\sqrt{2\pi}} 
          \left( \begin{array}{c} 
           e^{ i\frac{\theta}{2} } \hat{f}_{+,\mu,E}(x) \\  
           e^{-i\frac{\theta}{2} } \hat{f}_{-,\mu,E}(x) \end{array} 
              \right) \; ,
\ee
the same normalization condition as in Ref. \onlinecite{bardeen}.
The function satisfies the Bogoliubov-de Gennes equation, Eq. (30),
\bea
  & &  \frac{\hbar^2}{2m} \left[ - \frac{d^2 }{dr^2} 
    - \frac{1}{r}\frac{ d }{d r} + \frac{(\mu + \frac{1}{2})^2}{r^2} 
      - k_{\rho}^2 \right] \hat{f}_{+,\mu,E}(x)  \nonumber \\
  & &    + |\Delta(r)| \hat{f}_{-,\mu,E}(x) = E  \hat{f}_{+,\mu,E}(x) \; 
\eea
and 
\bea
 & & -\frac{\hbar^2}{2m} \left[ - \frac{d^2 }{dr^2} 
     - \frac{1}{r}\frac{ d }{d r} + \frac{(\mu - \frac{1}{2})^2}{r^2} 
      - k_{\rho}^2 \right] \hat{f}_{-,\mu,E}(x) \nonumber \\
 & & + |\Delta(r)| \hat{f}_{+,\mu,E}(x) = E  \hat{f}_{-,\mu,E}(x) \; .
\eea
Here $ r = | x- x_0|$ and  $ k_z^2 + k_{\rho}^2 = k_F^2 $.    
Inside the vortex core, we may set the energy gap to zero, $ |\Delta(r)| = 0$.
There are two independent solutions in this region, which we may choose
to be the following forms:
\be 
   \hat{f}_{1,\mu,E}(x) =  \frac{1}{ \sqrt{2}} 
       \left( \begin{array}{c}  1  \\ 0  \end{array} \right)
     J_{ \mu + \frac{1}{2} }
      \left(\sqrt{ k_{\rho}^2 + 2 m |E| /\hbar^2 } \; r\right)\; 
\ee
and 
\be 
   \hat{f}_{2,\mu,E}(x) =  \frac{1}{\sqrt{2}} 
       \left( \begin{array}{c}  0 \\ 1 \end{array} \right)
    J_{ \mu - \frac{1}{2} }
     \left(\sqrt{ k_{\rho}^2 - 2 m |E| /\hbar^2 } \; r\right) \; .
\ee 
Here $J_{ \mu \pm \frac{1}{2} } $ are Bessel functions.    
Away from the zero-energy gap region, 
the corresponding solutions may take the forms
\bea 
   & & \hat{f}_{1,\mu,E}(x) \nonumber \\
   & & =  \frac{1}{\sqrt{2}} 
       \left( \begin{array}{c} 
    \sqrt{ 1 + { \sqrt{ E^2 - |\Delta|^2 } }/{E} }  \\ 
    \sqrt{ 1 - { \sqrt{ E^2 - |\Delta|^2 } }/{E} }  \end{array} \right)
        J_{ \mu + \frac{1}{2} }\left( k_+(E) r\right)   \; ,
\eea
and 
\bea 
   & & \hat{f}_{2,\mu,E}(x)  \nonumber \\
   & & =  \frac{1}{\sqrt{2}} 
       \left( \begin{array}{c} 
    \sqrt{ 1 - { \sqrt{ E^2 - |\Delta|^2 } }/{E} }  \\ 
    \sqrt{ 1 + { \sqrt{ E^2 - |\Delta|^2 } }/{E} }  \end{array} \right)
        J_{ \mu - \frac{1}{2} }\left( k_-(E) r\right)   \; ,
\eea
where $ k_{\pm}(E) = \sqrt{ k_{\rho}^2 
  \pm 2 m \sqrt{ E^2 - |\Delta|^2 }  /\hbar^2 } $.
One may check that Eqs. (A6) and (A7) give the asymptotically exact solutions
when $r = \infty$.
They are WKB-type solutions connected to the solutions at
$r=0$ and $r = \infty$, valid under the condition that 
the energy gap $|\Delta|$ is smooth on the scale of $1/k_F$.
Exact solutions may be difficult to find. However, for the present purpose of 
demonstration of the diverging overlap integrals they are good enough.  
The solutions for a negative energy $-E$ can be constructed by using Eq. (37):
\bea 
   & & \hat{f}_{1,-\mu,-E}(x) \nonumber \\
   & &  =  \frac{1}{\sqrt{2}} 
       \left( \begin{array}{c} 
    \sqrt{ 1 - { \sqrt{ E^2 - |\Delta|^2 } }/{E} } \\ 
   -\sqrt{ 1 + { \sqrt{ E^2 - |\Delta|^2 } }/{E} } \end{array} \right)
        J_{ \mu + \frac{1}{2} }\left( k_+(E) r\right)   \; 
\eea
and 
\bea 
   & & \hat{f}_{2,-\mu,-E}(x)  \nonumber \\ 
   & & =  \frac{1}{\sqrt{2}} 
       \left( \begin{array}{c} 
    \sqrt{ 1 + { \sqrt{ E^2 - |\Delta|^2 } }/{E} }  \\ 
   -\sqrt{ 1 - { \sqrt{ E^2 - |\Delta|^2 } }/{E} } \end{array} \right)
        J_{ \mu - \frac{1}{2} }\left( k_-(E) r\right)   \; .
\eea
The immediate conclusion of the thermodynamic limit
is that there is an infinite degeneracy for a given energy
characterized by $\mu$, corresponding to the angular momenta of quasiparticles.
Those states form the base functions for the partial
wave analysis of the quasiparticle scattering, and make the transitions
between states with the same energy meaningful.

We now consider the left hand side of Eq. (38)
with an arbitrary small energy difference, 
\be
   {\bf I} \equiv 
   \int d^3 x \Psi^{\dag}_{\alpha}(x)  ( \nabla_{x_0}{\cal H}_0 )  
              \Psi_{\alpha'}(x) \; .
\ee
We will show that it can be a finite value (nonzero).
For the vanishing small trapping potential $U_0$, 
\be
   (\nabla_{x_{0}} {\cal H}_0 ) = 
           \left( \begin{array}{cc} 
                    0   & \nabla_{x_{0}} \Delta  \\
                   \nabla_{x_{0}}  \Delta^{\ast} & 0
            \end{array} \right) \; .
\ee
Since
\bea
   \nabla_{x_{0}} \Delta(x) & = & - e^{ i\theta} |\Delta(r)|'_{r}
           \,  \left(  \hat{x} \cos\theta + \hat{y} \sin\theta  \right)
            \nonumber \\
   & &    - i \, e^{ i\theta} |\Delta(r)| \,
           \frac{ - \hat{x} \sin\theta + \hat{y} \cos\theta }{ r } \; ,
\eea
the integral ${\bf I}$ may be expressed as
\bea
   {\bf I } & = & \delta_{k_z,k_z'} \int_0^{\infty} r dr 
                    \int_0^{2\pi} \frac{d \theta }{2\pi} 
                                      e^{- i (\mu-\mu' )\theta }  
           \left[ \hat{f}_{+,\mu,E}^{\ast}(r) \times \right. \nonumber \\
   & &  \left.  \nabla_{x_0} \Delta  \hat{f}_{-,\mu',E'}(r)
     +  \hat{f}_{-,\mu,E}^{\ast}(r) 
     \nabla_{x_0}\Delta^{\ast} \hat{f}_{+,\mu',E'} (r) \right] \nonumber  \\
          & = & \frac{1}{2} \delta_{k_z,k_z'} \, \delta_{\mu', \mu\pm 1}
               \left[ \hat{x} (a_{\mu,\mu'}(E,E') \pm b_{\mu,\mu'}(E,E') )  
        \right. \nonumber \\
   & &  \left. + i \hat{y} (\pm a_{\mu,\mu'}(E,E') 
          + b_{\mu,\mu'}(E,E')) \right] \; ,
\eea
where $\hat{x}$($\hat{y}$) is the unit vector in the $x(y)$ direction,
\bea
   a_{\mu,\mu'}(E,E') & = & - \int_0^{\infty} r dr |\Delta|'_r 
        \left[ \hat{f}_{+,\mu,E}^{\ast}(r) \hat{f}_{-,\mu',E'}(r)
          \right. \nonumber \\ 
   & & \left. + \hat{f}_{-,\mu,E}^{\ast}(r) \hat{f}_{+,\mu',E'}(r) \right] \; 
\eea
and 
\bea
   b_{\mu,\mu'}(E,E') & = & - \int_0^{\infty}  dr |\Delta|  
     \left[ \hat{f}_{+,\mu,E}^{\ast}(r) \hat{f}_{-,\mu',E'}(r) \right. 
         \nonumber \\
   & & \left. - \hat{f}_{-,\mu,E}^{\ast}(r) \hat{f}_{+,\mu',E'}(r) \right] \; .
\eea
The $\cos\theta$ and $\sin\theta$ inside the integral
give rise to the selection rule for the transition elements:
$ \int_{0}^{2\pi} d\theta  e^{ - i(\mu -\mu')\theta } \, \cos\theta 
   =  \pi \, \delta_{\mu', \mu\pm 1} $ and
$ \int_{0}^{2\pi} d\theta  e^{ - i(\mu -\mu')\theta } \, \sin\theta 
   = \pm i \pi \, \delta_{\mu', \mu\pm 1} $.
Only the transition between neighboring $\mu's$, that is, $\mu' = \mu \pm 1$,
can be nonzero.

For those nonzero $a_{\mu,\mu \pm 1}(E,E')$,
the only possible place which may give rise to an infinite value
for the integral is the region far away from the vortex core.
In this region, the eigenfunctions are given in the form of Bessel 
functions, Eqs. (A6)-(A(9) (c.f. Eq. (4.10) of Ref.\onlinecite{bardeen}), 
which are well behaved.
Away from the core $|\Delta(r)|'_{r}$ goes to zero rapidly, 
we conclude that the integral $a_{\mu,\mu'}(E,E')$ 
containing $|\Delta(r)|'_r$ is finite.
It may be instructive to give an estimation of $a_{\mu,\mu \pm 1}(E,E')$.
For this purpose we consider $\mu > 0$ and $\mu' = \mu + 1$ for 
positive energy states.
The negative $\mu$ and negative energy cases give similar results.
We will also restrict to the case that 
$E$ and $E' $ are close to each other,  
i.e., $\left| E - E' \right| < \Delta_\infty$,
and that both are sufficiently close to the energy gap, 
i.e., $E \sim \Delta_\infty$.
In this case the integral has  the largest value, and it incorporates the equal
energy limit implied in Eqs. (47) and (49).
First, we note that 
since  $k_\pm (E) \approx k_F$ and 
$J_\mu(k\pm(E) r) =  (k_\pm(E) r)^\mu/\mu!$  for small $k(E) r$,
the Bessel function is negligible small if $  r < r_t = \mu/k_F$ and
$\mu$ is large. The integral for region $r < r_t$ is negligible.
Second, we also neglect the integral in the region $r > \xi_0$, because
$|\Delta(r)|'_r$ is small. 
The integral now becomes
\be
   a_{l,\mu} \approx  \left\{
         \begin{array}{lcl} - \int_{r_t} ^{\xi_0} r dr |\Delta|'_r 
        \left[ \hat{f}_{+,\mu,E}^{\ast}(r) \hat{f}_{-,\mu+1,E'}(r) \right.
                & & \\
       \left. {\ } {\ }     
            + \hat{f}_{-,\mu,E}^{\ast}(r) \hat{f}_{+,\mu+1,E'}(r) \right]
         & , & r_t < \xi_0    \\
       0 & , & r_t > \xi_0 \; . \end{array} \right.   
\ee
Here we have used $l$ to denote the various combinations
from the solutions, Eqs. (A6)-(A9), specified below.
In the region $r < \xi_0$, $|\Delta|'_r \approx \Delta_\infty/\xi_0$
in the above integral. 
For $k_+(E),k_+(E')$ ($k_-(E),k_-(E')$ can be considered in 
the same manner.), following Eqs. (A16) and (A6)  we have
\bea
   a_{1,\mu} & = & 
         - \int_{r_t} ^{\xi_0} r dr \frac{\Delta_\infty}{\xi_0} 
             \times \nonumber \\
     & & 
       \frac{1}{2}
  \left[ \sqrt{ \left( 1 + \frac{ \sqrt{ E^2 - |\Delta|^2 } }{E} \right)
                \left( 1 - \frac{ \sqrt{ E'^2 - |\Delta|^2 } }{E'} \right) }
         \right.   \nonumber \\ 
   & & \left. 
   + \sqrt{ \left( 1 - \frac{ \sqrt{ E^2  - |\Delta|^2 } }{E} \right)
            \left( 1 + \frac{ \sqrt{ E'^2 - |\Delta|^2 } }{E'}\right) } \right]
       \times  \nonumber \\
   & &  J_{\mu + \frac{1}{2}}\left( k_+(E) r\right) 
        J_{\mu + \frac{3}{2}}\left( k_+(E') r\right)   \; .
\eea
Using the asymptotic form of Bessel function,
$ J_{\nu }( z ) = \sqrt{ 2/\pi z}\, \cos ( z - \nu \pi/2 - \pi/4) $,
and approximating the factor in the square bracket by  2,
we find 
\bea
    a_{1,\mu} & \approx &  
         -  \frac{\Delta_\infty}{\xi_0} \frac{1}{2\pi k_F} 
         \int_{r_t} ^{\xi_0}  dr \sin[ \Delta k  \, r ] 
        \nonumber \\
     & =  & - \frac{\Delta_\infty}{2\pi k_F} \Delta k \xi_0 /2  \; .
\eea
Here $\Delta k \equiv k_+(E) - k_+(E') $.
In reaching Eq. (A18) we have dropped a smaller contribution 
from $\int_{r_t} ^{\xi_0}  dr \cos[ (k_+(E) + k_+(E'))r - (2\mu + 3)\pi/2]$, 
because $k_+(E) r_t = \mu > 1$, and have also used the fact that 
 $ |k_+(E) - k_+(E')| \xi_0 < 1 $.
Equation (18) gives $|a_{\mu,\mu+1}(E,E')| < \Delta_\infty /k_F$.

For $k_+(E),k_-(E')$, following Eqs. (A16), (A6) and (A7) we have
\bea
   a_{2,\mu} & = & 
      - \int_{r_t} ^{\xi_0} r dr \frac{\Delta_\infty}{\xi_0 } 
        \times \nonumber \\
      & &  \frac{1}{2}
  \left[ \sqrt{ \left( 1 + \frac{ \sqrt{ E^2 - |\Delta|^2 } }{E} \right)
                \left( 1 + \frac{ \sqrt{ E'^2 - |\Delta|^2 } }{E'} \right) }
         \right.   \nonumber \\ 
   & & \left. +
     \sqrt{ \left( 1 - \frac{ \sqrt{ E^2 - |\Delta|^2 } }{E} \right)
            \left( 1 - \frac{ \sqrt{ E'^2 - |\Delta|^2 } }{E'}\right) } \right]
       \times \nonumber \\
   & &   J_{\mu + \frac{1}{2}}\left( k_+(E) r\right) 
       J_{\mu + \frac{1}{2}}\left( k_-(E') r\right)   \; .
\eea 
Please note the difference between Eqs. (A17) and (A19) 
in the indices of the Bessel functions.
Since $k_+(E) - k_-(E') \approx 1/\xi_0$, and again 
approximating the factor in the square bracket by  2, 
we find that 
\bea
   a_{2,\mu} & = & \frac{\Delta_\infty}{\xi_0} \frac{1}{2\pi k_F} 
         \int_{r_t} ^{\xi_0}  dr \cos[ (k_+(E) - k_-(E'))r ] 
        \nonumber \\
            & \approx & - \frac{\Delta_\infty}{2\pi k_F} \; .
\eea

Now we consider the phase integral part of ${\bf I}$, 
the integral $b_{\mu,\mu'}(E,E')$.
For $ r=|x-x_0| \rightarrow \infty$, $|\Delta| \rightarrow \Delta_{\infty}$.
We may ignore the integral in the region $r < r_t$, 
but not when $r_t > \xi_0$.  
Keeping the leading contribution, $ b_{\mu,\mu'}(E,E')  $ may be expressed 
as 
\bea
    b_{l,\mu} & \approx &  - \Delta_{\infty}
      \int_{r_t}^{\infty}  dr 
     \left[ \hat{f}_{+,\mu,E}^{\ast}(r) \hat{f}_{-,\mu',E'}(r) \right. 
            \nonumber \\
   & & \left. - \hat{f}_{-,\mu,E}^{\ast}(r)\hat{f}_{+,\mu',E'}(r)\right] \;  .
\eea
The Bessel functions will be replaced by their
asymptotic forms inside Eq. (A21). 
In the following we consider four cases as done for $a_{\mu,\mu'}(E,E')$.
For $ k_+(E), k_+(E')$, following Eqs. (A21) and (A6) we have
\bea
   b_{1,\mu} & = & 
         - \Delta_\infty  \int_{r_t} ^{\infty} dr \nonumber \\
      & &  \frac{1}{2}
  \left[ \sqrt{ \left( 1 + \frac{ \sqrt{ E^2 - |\Delta|^2 } }{E} \right)
                \left( 1 - \frac{ \sqrt{ E'^2 -|\Delta|^2 } }{E'}\right) }
         \right.    \nonumber \\ 
   & & \left.  
    - \sqrt{ \left(1 - \frac{ \sqrt{ E^2 - |\Delta|^2 } }{E} \right)
             \left(1 + \frac{ \sqrt{ E'^2 - |\Delta|^2 }}{E'}\right)}\right]
       \times \nonumber \\
   & &  J_{\mu + \frac{1}{2}}\left( k_+(E) r\right) 
       J_{\mu + \frac{3}{2}}\left( k_+(E') r\right)   \; .
\eea 
Since $ E' \rightarrow E $ and both are close to the energy gap, 
the factor inside the square bracket is always 
an order of unity, and we approximate by 2.
However, we note that when $E=E'$, the term in the square bracket
approaches to zero when $r >> \xi_0$.
Using the asymptotic form of the Bessel functions, 
\bea
     b_{1,\mu}  & \approx &  - \Delta_\infty \int_{r_t} ^{\infty} dr 
     \frac{1}{2\pi k_F} \frac{ \sin[ \Delta k_{++} \, r ] }{r}
                       \nonumber \\ 
      & = &  - \frac{\Delta_\infty}{2 \pi k_F} 
            \left\{  \begin{array}{lcl} 
                 sgn(\Delta k ) \; {\pi}/{2}   
             & , & | \Delta k | r_t < 1 \\
      { 1}/{ \Delta k   r_t }
             & , & | \Delta k | r_t > 1 \end{array} \right.  
\eea
Here $\Delta k \equiv k_{+}(E) - k_{+}(E')$.
Again, the contribution from $ \int_{r_t} ^{\infty} dr  
{ \cos[ ( k_+(E) + k_+(E')) r - (2\mu + 3) \pi/2   ] }/{r} $ 
has been ignored, because $k_+(E) r_t > 1$.
Since $| k_+(E) - k_+(E')| \xi_0 < 1$, the condition 
$| k_+(E) - k_+(E')| r_t < 1 $ will be satisfied if $ r_t < \xi_0$.

For $ k_+(E), k_-(E')$, following Eqs. (A21), (A6), and (A7) we have
\bea
   b_{2,\mu} & = & 
         - \Delta_\infty  \int_{r_t} ^{\infty} dr \nonumber \\ 
    & &     \frac{1}{2}
  \left[ \sqrt{ \left( 1 + \frac{ \sqrt{ E^2 - |\Delta|^2 } }{E}  \right)
                \left( 1 + \frac{ \sqrt{ E'^2 - |\Delta|^2 } }{E'}\right ) }
         \right.     \nonumber \\ 
   & & \left. - 
     \sqrt{ \left( 1 - \frac{ \sqrt{ E^2 - |\Delta|^2 } }{E} \right)
            \left( 1 - \frac{ \sqrt{ E'^2 - |\Delta|^2 }}{E'}\right)} \right]
      \times \nonumber \\
    & & J_{\mu + \frac{1}{2}}\left( k_+(E) r\right) 
       J_{\mu + \frac{1}{2}}\left( k_-(E') r\right)   \; .
\eea 
Using a similar procedure for $b_{1,\mu}$ we find
\bea
    b_{2,\mu} & \approx &  - \Delta_\infty \int_{r_t} ^{\infty} dr 
     \frac{1}{2\pi k_F} \frac{ \cos[ ( k_+(E) - k_-(E')) r ] }{r}
                        \nonumber \\   
     & = &  - \frac{\Delta_\infty}{2 \pi k_F} 
         \left\{  \begin{array}{lcl}    
              O(1)  & , &  r_t/ \xi_0 \; < 1 \\
                          { \xi_0 }/{ r_t }
                & , &  r_t/\xi_0  > 1 \; . \end{array} \right.  
\eea
We have used $ k_+(E) - k_-(E') \approx 1/\xi_0 $ in Eq. (A29).
If one is concerned about the logarithmic divergence of the cosine integral 
when  $ k_+(E) - k_-(E') \rightarrow 0$, we point out that it only occurs
when both  $|E|$ and $|E'|$ are approaching the energy gap $\Delta_\infty$.
In this limit the factor inside the square bracket goes to zero linearly,
and completely removes the logarithmic factor from the cosine integral.

The conclusion which one may draw from Eqs. (A23) and (A25) is  
that the integral $ b_{\mu,\mu' \pm 1}(E,E')$ is finite.
Together with what we have obtained for 
$a_{\mu,\mu' \pm 1}(E,E')$, 
the integral ${\bf I}$, therefore, the left side of Eq. (38), is finite.

Using Eq. (38), we have the overlap integral,
\be
  {\bf II } \equiv \int d^3 x \Psi^{\dag}_{\alpha}\nabla_{x_0} 
               \Psi_{\alpha'}(x)
            = \frac{\bf I}{E_{\alpha'} - E_{\alpha } }  \; . 
\ee
Since ${\bf I}$ is finite for the case of $\mu' = \mu \pm 1$
when $ E_{\alpha'} - E_{\alpha} \rightarrow 0$, ${\bf II}$
diverges as $1/(E_{\alpha'} - E_{\alpha} )$. 
Because asymptotically from the vortex core 
the wave functions $\Psi_\alpha$ always approaches a Bessel function,
this diverging behavior may be  directly
deduced from the right-hand side of Eq. (38) 
with the  aid of the recurrence relations of the Bessel functions.
The advantage of the demonstration here is that the right-hand
side of Eq. (49) is finite when two energies are exactly equal, without
the explicit consideration of the diverging behavior of the overlap integral.

There are two comments worthwhile to  make.   

1.  The existence of the limit at the left-hand side of Eq. (38)
    in the zero-energy difference indicates that the spectral function
    of Eq. (49) is a smooth function for small frequencies, and 
     it may be characterized by a power of the frequency.    

2.  This limiting behavior also removes the paradox that
    the frictional coefficient involves inelastic processes, but
    it may be obtained by calculating the elastic-scattering crossection of 
    quasiparticles implied in the thermodynamic limiting procedure.

\end{document}